\documentclass[preprint,epsfig]{aastex}
\usepackage{times}
\usepackage{graphicx}        

\begin{document}

\title{Discovery of black hole spindown in the BATSE catalogue of long GRBs}
\author{Maurice H.P.M. van Putten}
\affil{Korea Institute for Advanced Study, Dungdaemun-gu, Seoul 130-722, Korea \email{mvputten@kias.re.kr}}
\begin{abstract}
The BATSE catalogue is searched for evidence of spindown of black holes or proto-neutron stars (PNS) by extracting normalized light curves (nLC). The nLC are obtained by matched filtering, to suppress intermediate time scales such as due to the shock break-out of GRB jets through a remnant stellar envelope. We find consistency within a few percent of the nLC and the model template for spindown of an initially extremal black hole against high-density matter at the ISCO. The large BATSE size enables a study of the nLC as a function of durations $T_{90}$. The resulting $\chi^2_{red}$ is within a $2.35\sigma$ confidence interval for durations $T_{90}>20$ s, which compares favorably with the alternative of spindown against matter further out and spindown of a PNS, whose $\chi^2$ fits are, respectively, outside the 4$\sigma$ and 12$\sigma$ confidence intervals. {We attribute spindown against matter at the ISCO to cooling by gravitational-wave emission from non-axisymmetric instabilities in the inner disk or torus as the result of a Hopf bifurcation in response to energetic input from the central black hole.} This identification gives an attractive outlook for chirps in quasi-periodic gravitational waves lasting tens of seconds of interest to LIGO, Virgo and the LCGT.
\end{abstract}

\maketitle

\section{Introduction}

The nature of the inner engine of cosmological gamma-ray bursts has been an enigma ever since their serendipitous discovery by the Vela \citep{kle73} and Konus satellites, and even more so by their cosmological distribution and bi-modal distribution in durations established by the Burst and Transient Source Experiment (BATSE) \citep{kou93}. They are now known to take place about once per minute throughout the Universe, tracing the earliest galaxies known \citep{cuc11}. At least some of the long GRBs are associated with supernovae of relatively heavy stars \citep{pac91,woo93,pac98} following accurate pointing of their X-ray afterglow emission by Beppo-Sax \citep{pir95}, in a mass range that produces black holes or neutron stars. For a few long GRB events, we can rule out an accompanying supernova \citep{del06}, however, and these may, instead, be associated with merger events, e.g., of a neutron star with a companion black hole or another neutron star \citep{van08a}. This points to a common inner engine to long GRBs with or without supernovae, whose time scale of duration does {\em not} appear to depend on accretion. A secular time scale that extends to tens of seconds may arise in a process of spindown of black holes or neutron stars \citep{bis70,van99,bet03,van04,des07,des08}.

Some long GRB-supernovae are hyper-energetic \citep{cen10,cen11}, whose energy output exceeds the maximal rotational energy of a (proto-)neutron star (PNS) \citep{van11b}. This shows that PNS are not universal to all long GRBs which suggests, instead, that black holes are relevant for at least some of them. Distinguishing black holes and neutron stars as inner engines to long GRBs is particularly relevant for searches for gravitational waves associated with GRBs and core-collapse supernovae, in view of their rather distinct prospects for emission in gravitational radiation. Electromagnetic priors on the presence of black holes or neutron stars are hereby valuable in searches for gravitational-wave bursts by the advanced gravitational-wave detectors LIGO \citep{lig10}, Virgo and the Large Cryogenic Gravitational-wave Telescope (LCGT) \citep{lcgt10}. The spin energy of rapidly rotating stellar mass black holes exceeds the typical energy output as observed in the electromagnetic spectrum \citep{fra01} by some two orders of magnitude, thus providing ample opportunity to power substantial accompanying emission in MeV neutrinos and gravitational waves, that latter which pointing to a sensitivity distance of about 35 Mpc for a complete identification of a characteristic quasi-periodic chirp by a time sliced matched filtering (TMSF) method \citep{van11}. For long GRBs from mergers, an additional long duration radio bursts emanating from magnetic torus winds may be searched for by wide area radio survey instruments such as LOFAR \citep{van09b}. 

The prompt GRB emissions derive from dissipation in ultra-relativistic baryon-poor jets (BPJ) directed close to the line of sight to the observer, that hereby tracks the evolution of a long-lived inner engine by causality. The BPJ may emanate from magnetic outflows from the event horizon of a black hole \citep{lev93} or from a new born proto-neutron star (PNS, e.g. \cite{des07,des08}). The prompt GRB emission is hereby representative for the light curve emanating from the black hole or PNS. For black holes, these outflows can arise as capillary jets powered by frame dragging, provided that the black hole is rotating and intermittently active \citep{van08b}. In general terms, frame dragging introduces an energetic interaction between the angular momentum of particle outflow aligned with the angular momentum of the black hole. Frame dragging has recently been measured by LAGEOS satellites \citep{ciu04} and Gravity Probe B  \citep{eve11}. Frame dragging induced outflows hereby establish a direct link between gamma-ray light curves and the evolution of black hole spin. It correctly predicts X-ray afterglows also from short GRBs \citep{van01a} and points to long GRBs as a common endpoint of core-collapse supernovae and mergers involving rapidly spinning black holes, that may unify long GRBs with and without supernovae, the halo event GRB070125 and the long event GRB 050911 with no X-ray afterglow. 

The long duration time scales in the light curves of long GRBs hereby reflect the secular behavior of the GRB inner engines, after scaling out the effect of redshifts in the observed durations and different lifetimes of the spin of black holes or PNS resulting from unknown magnetic field strengths. For black holes, we recall that these are scale free objects with essentially no memory of their astronomical progenitor systems, except for the initial mass and angular momentum. Black hole spindown is an alternative to spin up by accretion of matter \citep{bar70}, where spindown is against surrounding high density matter with accompanying low energy emissions that account for most of the energy output \citep{van99,van01a,van03,van08a}. The nLC is hereby sensitive to the state of the surrounding matter, in particular its location relative to the Inner Most Stable Circular Orbit (ISCO). Our focus, therefore, will be on the asymptotic behavior in durations longer than any of the (magneto-)hydrodynamical time scales that are not directly relevant to the evolution of the black hole such as, in core-collapse supernovae, the time scales associated with black hole formation or the time scale of shock break-out of jets plowing through a remnant stellar envelope. In the asymptotic range of long durations, therefore, we anticipate that the nLC of ensemble averages of light curves becomes meaningful in ways similar to the Phillips light curve of Type Ia supernovae \citep{phi93}.

The aforementioned phenomenology of long GRBs zooms in onto a specific inner engine of long GRBs: long GRBs with and others without supernovae point to black holes or PNS with a long duration time scale different from accretion, as in thin neutrino dominated accretion disk models \citep{pop99,che07} based on the assumption of negligible output from the black hole onto the surrounding matter. Instead, a long duration time scale may derive from spindown, for which some hyper-energetic events demonstrate that a PNS is not universally viable. It is therefore natural to search for a prevalence of spindown of black holes in the light curves of long GRBs, since the time-averaged behavior in light curves should track the secular evolution of the inner engine by causality.

For this purpose, we developed a novel time-domain analysis of the light curves of the complete BATSE catalogue long GRBs by application of matched filtering. The distribution of durations of long GRBs in the BATSE catalogue is rather broad, which goes beyond dilations due to their distribution in redshift, indicating an intrinsic spread in durations \citep{bal03}; confirmed by {\em Swift}, e.g., \cite{geh09}). Their distribution in durations. therefore, includes the diversity in intrinsic durations beyond redshift effects. The large number of 1491 long GRB events in the BATSE catalogue introduces an appreciable resolving power, that we shall exploit to extract an accurate normalized light curve (nLC) to differentiate between different models for spin down of black holes, with and without gravitational-wave emissions, and neutron stars. 

In \S2, we review the observed high energy emissions in GRBs. \S3 discusses a theory of spindown of black holes and PNS. \S4 presents three model templates for the resulting gamma-ray light curves in spindown of alternatively black holes, against matter at the ISCO or further out, or PNS. In \S5, we summarize the BATSE catalogue of light curves of long GRBs. The nLC of each model is given in \S6. We interpret our findings in \S7 for prolonged spindown of black holes differentiated for the state of matter surrounding the black hole, versus spindown of PNS with an outlook on long duration extragalactic bursts in radio and gravitational waves. Our findings are summarized in \S8.

\section{Long GRBs from rotating black holes or proto-neutron stars}

The GRB light curves represent broad band emissions in the gamma-rays. The Fermi/GLAST event GRB080916C \citep{abd09} reveals an extension of GRB spectra from 8 keV - 40 MeV of the Glast Burst Monitor (GBM) into the energy range 20 MeV - 300 GeV of the Large Area Telescope (LAT). GRB080916C is the most energetic event detected to date, of duration $T_{90}=66$ s, $z=4.25$, and with an isotropically equivalent energy output $E_{iso}=8.8\times 10^{54}$ erg with one additional photon detected at 13 GeV (70 GeV in its restframe). Here, $T_{90}$ refers to the time interval covering a 90 percentile in total photon count for a given burst \citep{kou93}. Its Band parameters $(E_0,\alpha,\beta)$ show temporal behavior around canonical values $(500\mbox{~keV},-1,-2.2)$ and imply a minimum Lorentz factor $\Gamma_j=890$ \citep{taj09}. The true energy in gamma-rays satisfies $E_\gamma=f_b^{-1}E_{iso}$, where $f_b=2/\theta_j^2$ denotes the beaming factor for a two-sided jet-like outflow with half-opening angle $\theta_j$. A number of GRB light curves display an achromatic break in their light curves, when the relativistic beaming angle $\sim 1/\Gamma_j$ of radiation produced by shock fronts exceeds the half-opening angle $\theta_j$ of the jet (e.g. \cite{mes06}). No clear measurement of $\theta_j$ has been made for GRB080916C, but $E_\gamma\sim 10^{52}$ erg is expected for the typical value $f_b\sim 500$. GRB080916C further reveals a delayed onset of maximal flux in the highest energy photons seen in LAT \citep{taj09}.

Prompt GRBs emissions have been associated with ultra-relativistic baryon poor jets emanating from black holes or neutron stars (e.g. \cite{tho94,met11}, where the former is an attractive alternative to account for low baryon-loading \citep{lev93,eic11}. 

If powered by the rotational energy of the black hole, they may produce essentially unlimited Eddington luminosities from stellar mass black holes for the duration of black hole spin. As capillary jets \citep{van08b} produced by frame dragging, they represent a direct link to the spin of the black hole.  (This process is distinct from the Penrose process, as it operates well outside the ergo sphere.) These capillary jets are non-uniform in luminosity and variability within the half-opening angle $\theta_j$ of the outflows (distinct from the half-opening angle $\theta_H$ on the black hole event horizon due to collimation, \cite{lev93,van03}), that reach their maximum at the interface with the environment or a surrounding baryon-rich disk or torus wind. Surfaces of maximal flux ($\theta\simeq \theta_j$) should hereby have maximal luminosity and variability, the latter by the expected turbulent boundary layer between the jet and a collimating wind \citep{van09}, consistent with the observed positive correlation between the two \citep{rei01}. The duration of the capillary jets is determined by the lifetime of black hole spin. Identified with the duration $T_{90}$ of long GRBs, it gives rise to a spectral-energy correlation $E_\gamma\propto E_pT_{90}^{1/2}$ between $E_\gamma$ and the peak energy $E_p$ in gamma-rays in agreement with observations by HETE II and Swift \citep{van08b} with a Pearson coefficient that is improved by including $T_{90}$.

The {\em Swift} event GRB060614 of duration $T_{90}=102$ s introduces a new class of long GRBs with no supernova and with a late-time X-ray tail (XRT), which generally extends on a timescale of 1 - 1000 ks common to long ($T_{90}>2$ s) and short GRBs ($T_{90}<2$ s) events alike \citep{zha07}. This suggests there is no apparent memory in the GRB remnant to the initial state giving rise to the prompt GRB emissions. The common endpoint of long and short GRB from rotating black holes are black holes with slow spin with no memory to initial spin. Subject to late time accretion \citep{ros07}, it would produce very similar XRTs to long and short events.

\section{Spindown of black holes and proto-neutron stars}

\begin{figure}[ht]
\centerline{\includegraphics[scale=.25]{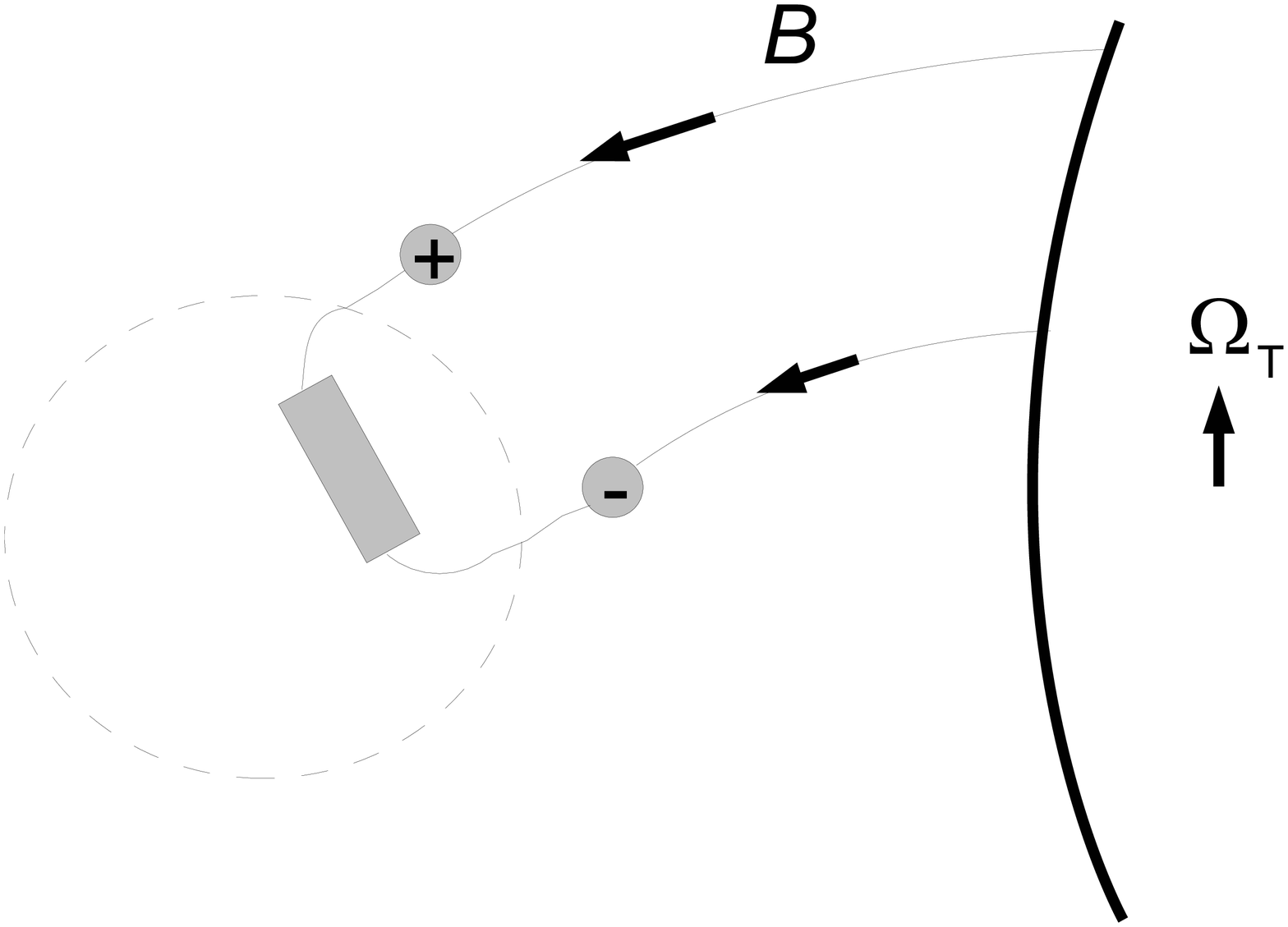}\includegraphics[scale=.25]{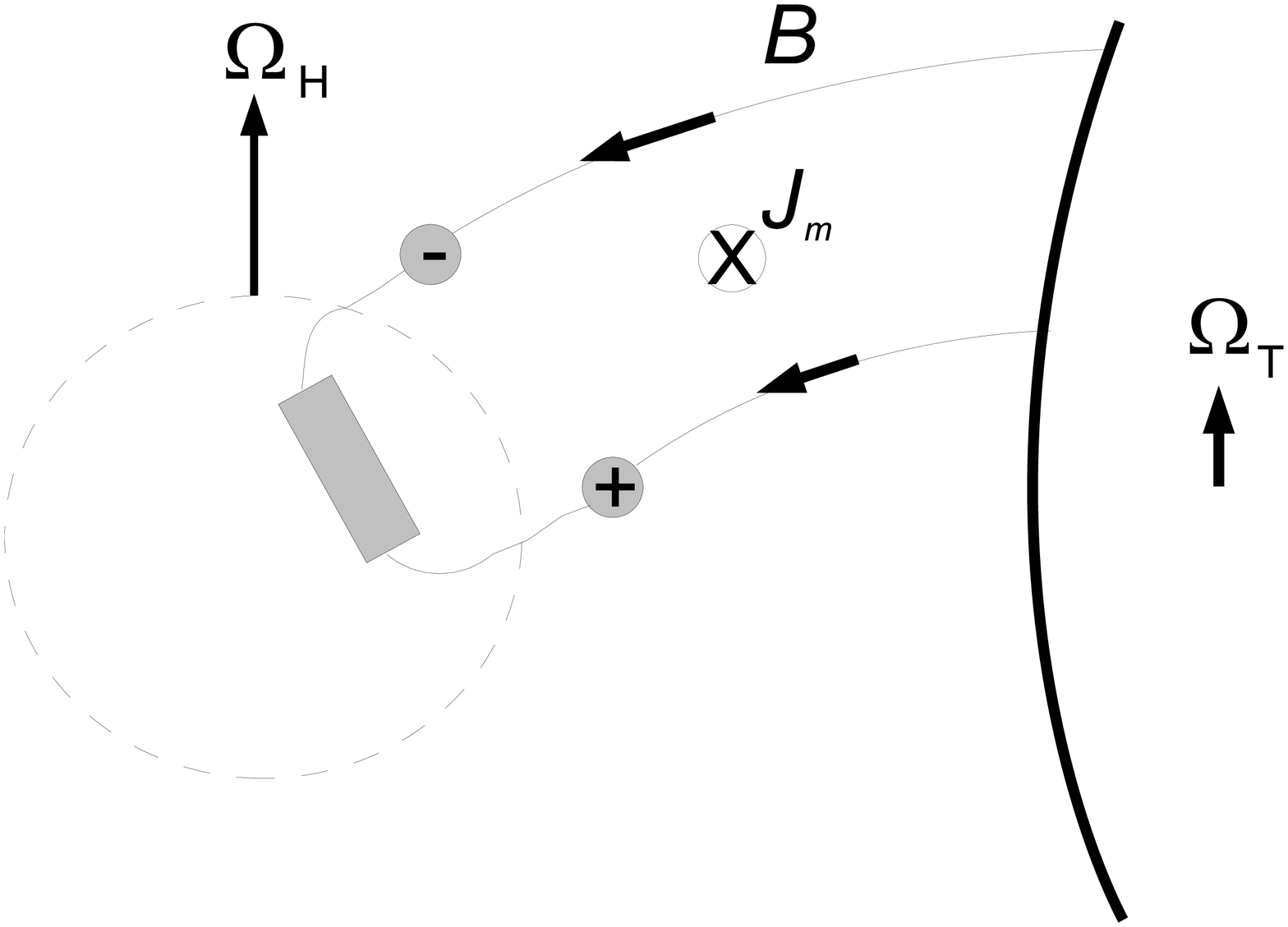}}
\caption{The interaction of frame dragging with magnetic fields is described by an additional Lie derivative ${\cal J}_m$ (\ref{EQN_JM}) in Faraday's equations (\ref{EQN_FA}). It induces poloidal current loops between rotating black holes and surrounding matter via an inner torus magnetosphere, here shown in poloidal cross-section with no-slip boundary conditions on the inner face of a torus and radiative (slip) boundary conditions on the event horizon of the black hole. The result is a potentially powerful spin-connection between the black hole (non-rotating, $left$; rotating, $right$) and the torus, wherein the black hole event horizon serves as a passive load. The induced poloidal currents mediate energy and angular momentum transfer by Maxwell stresses on the event horizon \citep{ruf75,bla77} and the inner face of the torus \index{torus} \citep{van99}. If the black hole spins rapidly, Lorentz forces due to the induced poloidal currents cause the inner face of the torus to spin up, equivalent to those in the magnetosphere of a spinning neutron star by which it spins down \citep{van03}.}
\label{FIG_MX1}    
\end{figure}

For black hole inner engines, we note that the spin energy of a Kerr black hole behaves remarkably similar to that of a spinning top in view of a ratio $\frac{E_{spin}}{\Omega_HJ}=\frac{1}{2}\cos^{-2}(\lambda/4)=0.5-0.5858$, where $\sin\lambda=a/M=J/M^2$ ($0\le \left|\lambda\right| \le \pi/2$), that remains remarkably close to the Newtonian limit of 1/2. Surrounding by high density matter, most of the black hole luminosity $L_H$ provided by $E_{spin}$ is expected to be incident onto the inner face of a torus by equivalence in poloidal topology to the magnetosphere of neutron stars, whereby its mass $M$ and spin evolves in response to a torque $T$ evolve according to \citep{van99}
\begin{eqnarray}
L_H=-\dot{M},~~T=-\dot{J}
\label{EQN_LM}
\end{eqnarray}
as a function of the black hole angular momentum $J$ and mass $M$. This interaction can also be seen by inspection of Faraday's equation for the electromagnetic field ($E,B$) as a consequence of frame dragging in Faraday's equations (e.g. \cite{tho86})
\begin{eqnarray}
\tilde{\nabla}\times \alpha {\bf E} = -\partial_t {\bf B} + 4\pi {\cal J}_m,
\label{EQN_FA}
\end{eqnarray} 
where $\alpha$ denotes the gravitational redshift and $\tilde{\nabla}_i$ denotes the 3-covariant derivative, and
\begin{eqnarray}
{\cal J}_m = -\frac{1}{4\pi} {\cal L}_\omega {\bf B}
\label{EQN_JM}
\end{eqnarray}
in terms of a Lie-derivative of the magnetic field with respect to $\omega$. The role of (\ref{EQN_JM}) in an inner torus
magnetosphere around a black hole is schematically illustrated in Fig. \ref{FIG_MX1}. 

A derivation of ${\cal J}_m$ based on the two Killing vectors of the Kerr metric is as follows. Let ${\bf F} = {\bf u}\wedge {\bf e} + * {\bf u} \wedge {\bf h}$ be the four-vector representation of the electromagnetic field $F_{ab}$ associated with $u^b$, $u^cu_c=-1$, of zero-angular momentum observers (ZAMOs) \citep{lic67}. Following \citep{sha83,tho86}, ${\bf u} = -\alpha {\bf d}t$ and ${\bf u}=\alpha^{-1}({\bf k}+\omega {\bf m})$ in the Killing vectors $k^b=(\partial_t)^b$ and $m^b=(\partial_\phi)^b$ of the Kerr metric. Consequently, $\nabla_cu^c=0$. ZAMOs measure $e^b=u_cF^{ac}$ and $h^b=u_c*F^{cb},$ where ${\bf e}=(0,E^i)$ and ${\bf h}=(0,B^i)$, where $*$ is the Hodge dual ($*^2=-1$).  To evaluate $\nabla_a*F^{ab}=0,$ we note 
\begin{eqnarray}
\nabla_a (u^ah^b-u^bh^a) = {\cal L}_u h^b + (\nabla_cu^c)h^b-(\nabla_ch^c)u^b,
\end{eqnarray}
where ${\cal L}_u h^b = (u^c\nabla_c)h^b - (h^c\nabla_c)u^b$. Projected onto surfaces of constant coordinate time $t$ (orthogonal to $u^b$), 
\begin{eqnarray}
\left({\cal L}_{\bf u} {\bf h}\right)_\perp = 
\alpha^{-1}\left(\partial_t {\bf B} + {\cal L}_{\omega^i} {\bf B}\right)
\end{eqnarray}
in the frame of ZAMOs, where $L_{\omega^i}$ is with respect to $\omega^i\equiv\omega m^i$ (where $m^i$ is not a unit three-vector).
Next, we decompose $\nabla_a = D_a- u_a(u^c\nabla_c),$ and note $*({\bf u}\wedge {\bf h})_{abcd} = \epsilon_{abcd}u^ce^d$, and the acceleration $(u^c\nabla_c)u_b=\alpha^{-1}\nabla_b\alpha$. Consider
\begin{eqnarray}
\nabla^b(\epsilon_{abcd}u^ce^d)=\epsilon_{abcd}(D^bu^c)e^d-\epsilon_{abcd}u^ba^ce^d
+\epsilon_{abcd}u^c\nabla^be^d.
\end{eqnarray}
The projection of the right hand side onto surfaces $(r,\theta,\phi)$ normal to $u^b$ satisfies $\epsilon_{ibcd}(D^bu^c)e^d+\tilde{\epsilon}_{ijk}a^je^k+\tilde{\epsilon}_{ijk}\nabla^je^k=\epsilon_{ibcd}(D^bu^c)e^d+\alpha^{-1}\tilde{\epsilon}_{ijk}\nabla^j(\alpha e^k),$ where $\epsilon_{aijk}u^a=\tilde{\epsilon}_{ijk}=\sqrt{h}\Delta_{ijk}$ with $\sqrt{-g}=\alpha\sqrt{h}$ over the 3-volume $\sqrt{h}$ of spacelike coordinates using $\Delta_{ijk}$, $\Delta_{123}=1$. Here, the first term on the right hand side vanishes, since $D_bu_c$ is spacelike: $u^b(D_bu_c)=0$ by construction and $u^cD_bu_c=0$ by normalization $u^2=-1$. The result is an extra current ${\cal L}_{\omega^i}B$ in Faraday's equation. See \cite{tho86} and references therein for an alternative derivation.

Applied to an inner torus magnetosphere \citep{van99} as in Fig. \ref{FIG_MX1}, we have
\begin{eqnarray}
\omega_i {\cal J}_m^i \simeq  \frac{1}{8\pi} {\bf B}\cdot \tilde{\nabla}(\omega_i\omega^i) > 0,~~
\omega_i\omega^i = 4\frac{z^2\sin^2\lambda}{(z^2+\sin^2\lambda)^3} ~~~\left(\theta=\frac{\pi}{2}\right),
\label{EQN_LOM}
\end{eqnarray}
where the inequality refers to an ingoing poloidal magnetic field with the orientation and sign of ${\cal J}_m$ as indicated. In this process, the event horizon serves as a passive load while torus, subject to powerful competing torques acting on the inner and the outer face of the torus, develops forced turbulence \citep{van99}. The complete evolution is now described by a system of two ordinary differential equations for the black hole luminosity $L_H$ and the torque $T_H$ by the induced Lorentz forces exerted (mostly) onto the surrounding matter \citep{van08a,van11}
\begin{eqnarray}
L_H=-\dot{M} = \kappa \left(\Omega_H-\Omega_T\right)\Omega_T,~~T_H=-\dot{J}=\kappa\left(\Omega_H-\Omega_T\right)
\label{EQN_MT}
\end{eqnarray}
parameterized by a spin down coefficient $\kappa$ given by the variance in poloidal magnetic field energy, the angular velocity $\Omega_T$ of the torus and $\Omega_H$ of the black hole in  their lowest energy state. Here, the angular velocity of the inner torus magnetosphere is equal to $\Omega_T$ by no-slip boundary conditions on the torus and slip boundary conditions on the event horizon of the black hole. This output $(L_H,T_H)$ will be re-radiated in various emission channels by the inner disk or torus in a catalytic process, mostly so in magnetic winds, MeV-neutrinos and gravitational waves.

In modeling long GRBs from rapidly rotating black holes, the open magnetic flux tube is supported by an equilibrium magnetic moment of the black hole \citep{wal74,van01b} whenever the black hole is exposed to an external poloidal magnetic field. We propose a positive correlation between $\theta_H$ and the radius of the torus described by a jet luminosity \citep{van09}
\begin{eqnarray}
 L_j \propto \Omega_H^2 z^n {\cal E}_k,
 \label{EQN_j}
\end{eqnarray}
where $R_T=zM$ for a black hole of mass $M$, ${\cal E}_k\propto (\Omega_TR_T)^2e(z)$ is the kinetic energy in the torus, and $e(z)=\sqrt{1-2/3/z}$ denotes the specific energy of matter in an orbit with angular velocity $\Omega_T$ at the ISCO \citep{bar70}. With $n=2$, $L_j$ scales with the surface area enclosed by the torus, while the scaling to ${\cal E}_k$ follows from a stability limit on the poloidal magnetic flux energy that the torus can support \citep{van03}. The limited half-opening angle of $\theta_H\simeq 0.15$ rad from the event horizon accounts for the fraction of about 0.1\% in true energies in gamma-rays relative to the spin energy of the black hole \citep{van03}.

For PNS, we recall that the evolution of their angular velocity with spin aligned magnetic field derives from a spindown luminosity (e.g. \cite{kal11}) 
\begin{eqnarray}
L_{PNS} \propto \Omega^4
\label{EQN_LPNS}
\end{eqnarray}
as a function of its angular velocity $\Omega$. Here, we assume that gravitational-wave emissions are negligible, which otherwise might derive, with considerable uncertainty, from a 
variety of dynamical processes (e.g. \cite{ree74,owe98,cut02,how04,and11,reg11}). The rotational energy, luminosity and spindown time for an initially maximally rotating PNS are
\begin{eqnarray}
E_{rot}\simeq E_c,~~L_c\simeq 10^{51}B_{16}^2 \left(\frac{M}{1.45M_\odot}\right)^2~\mbox{erg~s}^{-1},~~T_c\simeq \frac{E_c}{L_c}= 30B_{16}^{-2}~\mbox{s}
\label{EQN_Ec}
\end{eqnarray}
for a magnetic field strength $B=B_{16}10^{16}$ erg, where 
\begin{eqnarray}
E_c=3\times 10^{52}\mbox{~erg}
\end{eqnarray}
where $E_c=3\times 10^{52}\mbox{~erg}$ denotes the maximal rotational energy of a PNS. The luminosity $L_{PNS}$ is radiated off in a magnetic wind that may serve to drive an explosion and especially so following core-collapse of relatively heavy progenitor stars, while the possibility of it driving a GRB \citep{met11} depends crucially on baryon loading that may be prohibitive during the initial hot phase of the PNS \citep{des07,des08}. 

\section{Template light curves for spindown of black holes and proto-neutron stars}

The light curves of long GRBs show considerable diversity in durations, count rates, and variability, wherein uncertainties in photon count rates are relatively small given the large count rates that are often observed. Our focus is on extracting information on the secular (slow time) evolution of the inner engine as described in the previous section in the sense of an ensemble average. We do not intent to model detailed behavior in any individual burst as it may arise from, e.g., intermittencies in any of the (magneto-)hydrodynamic and radiation processes. Correspondingly, we set out to normalize the light curves of long GRBs to unit duration and integrated photon counts, and average these to extract a normalized light curve (nLC) \citep{van09}.

In the present study, we consider light curve model templates for GRBs from spindown of black holes by (\ref{EQN_j}) upon solving numerically for (\ref{EQN_MT}) with the closure relations for $\Omega_T$ given by
\begin{eqnarray}
\mbox{Model A}:~~\Omega_T=\Omega_{ISCO},~~\\
\mbox{Model~B}:~~\Omega_T=\frac{1}{2}\Omega_H,~~~~~
\label{EQN_MA}
\end{eqnarray}
Model A considers the torus to be at the ISCO, where $\Omega_{ISCO}=\pm \frac{1}{z^{3/2}\pm \sin\lambda}$ for co-rotating (+) and counter-rotating (-) orbits, where $z=r/M$ \citep{sha83}. For an initially maximally rotating black hole, the ISCO (+) expands in size from $M$ to $6M$ in the Schwarzschild limit of zero rotation for co-rotating orbits. Model B considers the torus to be further out. The templates are calculated by numerical integration of (\ref{EQN_MT}) with closures (\ref{EQN_MA}) with an initially maximal spin. The minor energy output in gamma-rays (\ref{EQN_j}) are modeled subsequently as a function of the calculated evolution of the black hole. 

\begin{center}
\begin{figure}[ht]
\center{\includegraphics[angle=00,scale=.65]{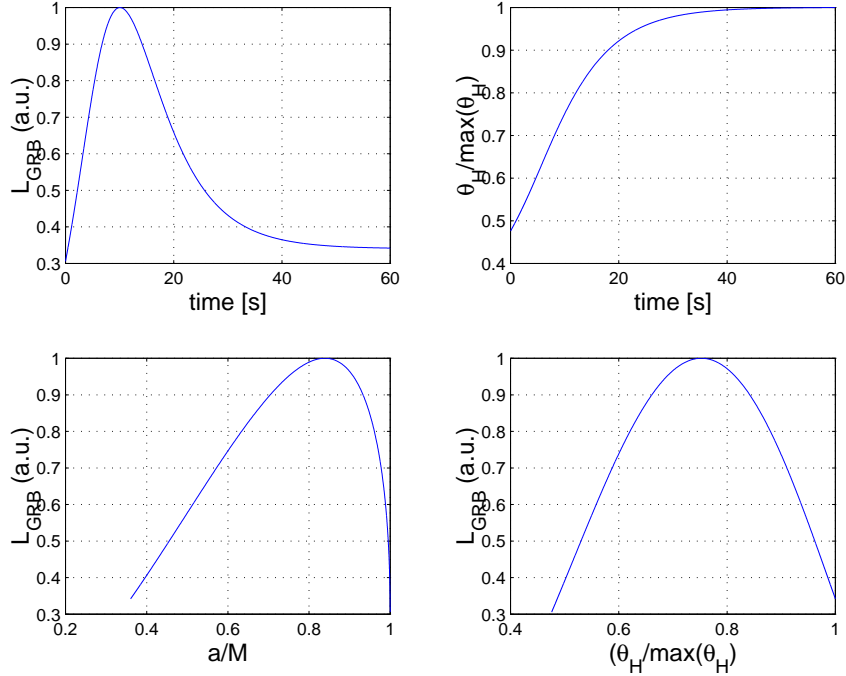}}
\caption{Shown is the template A for the light curve in gamma-rays for an initially extremal black hole and the associated horizon half-opening angle $\theta_H$ following (\ref{EQN_j}), representing spindown of a black hole against matter at the ISCO. The luminosity initially rises to maximal luminosity as $\theta_H$ increases with the expansion of the ISCO. At high spin rates $0.8388<a/M<1$, therefore, there exists an anti-correlation of luminosity with black hole spin. Template A closely matches relatively smooth light curves such as those of GRB 970508 and GRB 980425 (Fig. 1 in \cite{van09}).}
\label{FIG_T}
\end{figure}
\end{center}

On the basis of (\ref{EQN_j}),(\ref{EQN_MT}) and (\ref{EQN_MA}), we calculate a model light curve $L_j(t)$ of the minor output in high energy emissions, to represent the intrinsic light curve in gamma-rays detailed in \cite{van09}. Fig. \ref{FIG_T} shows the result for closure A, where $L_j(t)$ starts at a finite value $L_j(t_0)>0$ at the time of onset $t_0$ for an initially maximally spinning black hole, and gradually increases to a maximum when $a/M=0.8388$ before decaying to a finite value as $\Omega_H$ approaches $\Omega_{ISCO}$. For an initially extremal black hole, the maximum is attained with a delay
\begin{eqnarray}
\left(\frac{\Delta t}{T_{90}}\right)_A\simeq 16\%
\label{EQN_tau}
\end{eqnarray}
relative to $T_{90}$ of the model burst with $L_j(t)-L_j(t_0)\ge0$, where max $L_j(t)/L_j(t_0)=2.27$. With closure B, $L_j(t)$ starts promptly at near-maximum, and rapidly decays with black hole spin. For an initially maximally spinning black hole, numerical integration of (\ref{EQN_MT}) shows an overall efficiency of close to 60\% (equal to when $a/M=0.8$ initially) for A and an overall efficiency of 35\% for B, as follows by direct integration of (\ref{EQN_MT}) subject to the two alternatives (\ref{EQN_MA}). These results point to dissipation of a major fraction of black hole spin energy ``unseen" in the event horizon, creating astronomical amounts of entropy \citep{bek73}. Black hole spin down by (\ref{EQN_MT}) is largely {\em viscous.} 

As shown in Fig. \ref{FIG_T}, the luminosity in template A is initially anti-correlated to black hole spin while $a/M\simeq 0.8388$, associated with the increase of $\theta_H$ due to the expansion of the ISCO. This anti-correlation is consistent with a recent observation of no or a weakly negative correlation between jet luminosity and spin rates that are high \citep{fen10}.

A light curve template for GRBs from PNS may be derived from their spindown luminosity (\ref{EQN_LPNS}). Characteristic values for their rotational energy, luminosity and spindown time are those of (\ref{EQN_Ec}). We thus have
\begin{eqnarray}
\mbox{Model~C:}~L_j(t)=\frac{L_0}{(1+t/T_c)^2}
\label{EQN_PNS}
\end{eqnarray}
upon neglecting any gravitational wave emissions as mentioned above.

In comparing model light curves A-C above with the light curves in the BATSE 4B catalogue in the time domain, we shall assume that photon count rate integrated across the four BATSE energy bands is proportional to the luminosity $L_j(t)$. This assumption refers to the temporal behavior in any given burst, not across different bursts that further involves a diversity in parameters, e.g., baryon loading, that may affect the efficiency in prompt GRB emission processes. The template for Model A is shown in Fig. \ref{FIG_T}.

For the purpose of searching for evidence of spindown of a black hole or neutron star in long GRBs, we set out to consider long GRBs {\em on average}. To do so, we shall exploit BATSE catalogue of long GRBs.

\section{Matched filtering applied to the BATSE catalogue of 1491 long duration bursts}

BATSE produced by far the largest catalogue of GRB light curves. Since there is no report of any degradation of the BATSE instrument with time, this method ensures a completely unbiased and blind selection of events. It reveals a bi-modal distribution of short and long bursts, represented by $T_{90} < 2$ s and $T_{90} > 20$ s, respectively, as summarized in Fig. \ref{FIG_match01}.
\begin{center}
\begin{figure}[ht]
\center{\includegraphics[scale=.6]{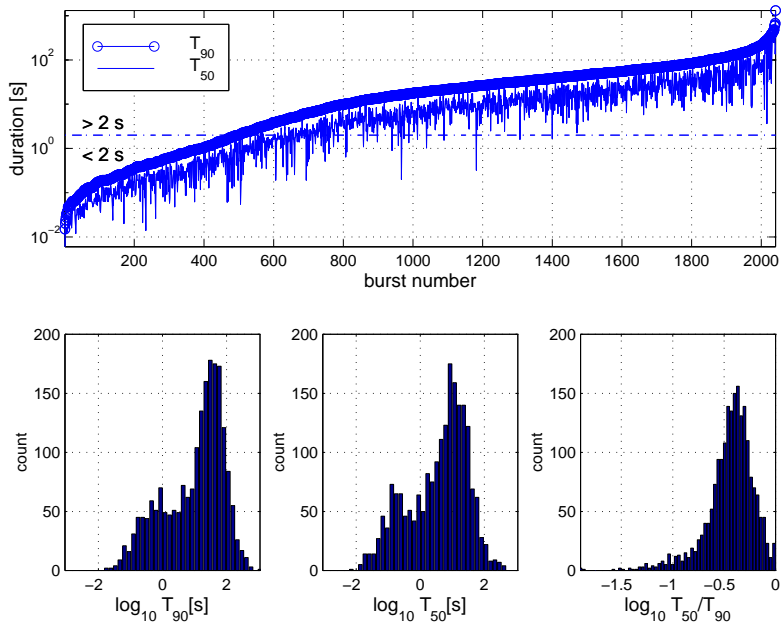}}
\caption{($Top.$) Overview of the durations $T_{90}$ and $T_{50}$ in the 1491 long GRBs in the BATSE Catalogue, here sorted by $T_{90}$. ($Bottom.$) The distribution is bi-modal in both durations, while the $\log_{10} T_{50}/T_{90}$ ratio (with a mean of 0.407) is subject to considerable scatter due to variability in the GRB light curves.}
\label{FIG_match00}
\end{figure}
\end{center} 

The $1491$ light curves  of long GRBs comprise 531 and 960 events with durations 2 s $<T_{90}<$ 20 s and $T_{90}>$ 20 s, respectively. 
For the purpose of our study, we first consider the normalized light curves (nlc$_i$) of each burst ($i=1,2,\cdots N$) individually as a function of normalized time $\tau = t/T_{90}$ in the interval
\begin{eqnarray}
I= [-1,3],
\end{eqnarray}
subsequent to filtering out high frequency fluctuations by applying a moving averaging over 40 samples over 64 ms (corresponding to 2.56 seconds) to the BATSE light curve data. For the BATSE catalogue of 1491 light curves with $T_{90} > 20$ s, the results are summarized in Figs. \ref{FIG_match01}.
\begin{figure}[ht]
\centerline{\includegraphics[scale=.85,angle=00]{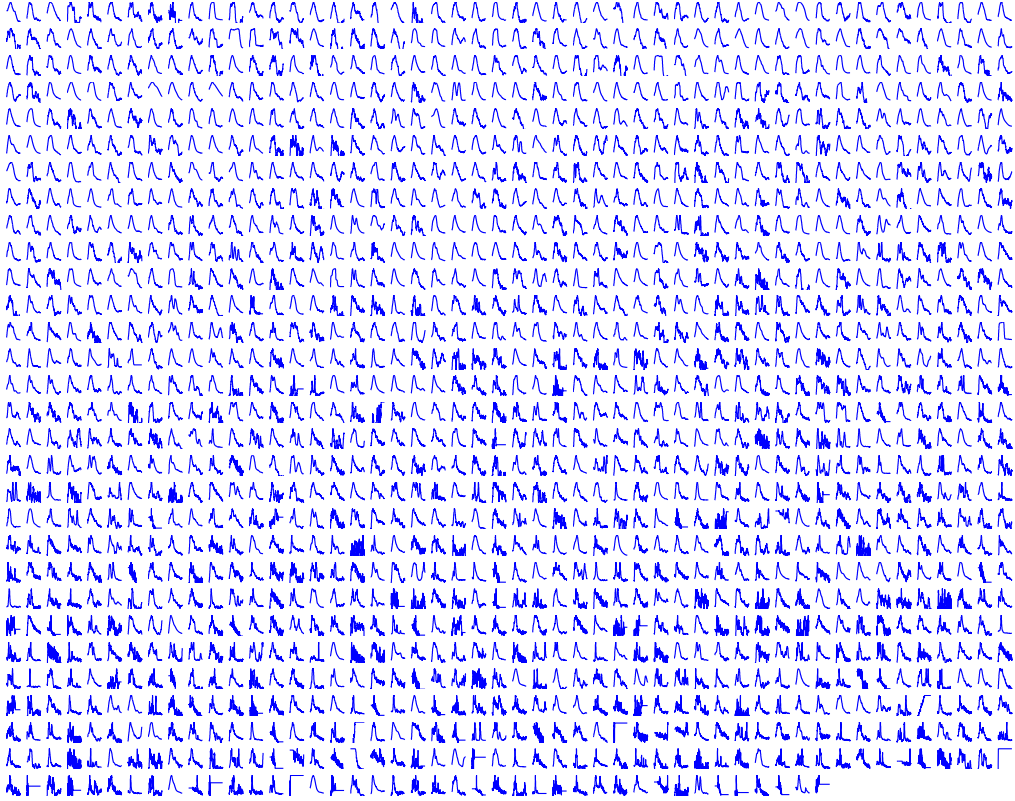}} 
\caption{Compilation of the complete BATSE catalogue of 1491 light curves of long bursts sorted by 2 s $< T_{90}\le  $1307 s.
Each light curve shown represents the sum of the photon count rate in all 4 BATSE energy channels, is smoothed with a time scale of 2.56 seconds and is plotted as a function of time normalized to $T_{90}$ in the interval of $[-1,3]$. We note an apparent trend from relatively similar to increasingly diverse light curves with $T_{90}$.}
\label{FIG_match01}    
\end{figure}

\begin{figure}[ht]
\centerline{\includegraphics[scale=.7]{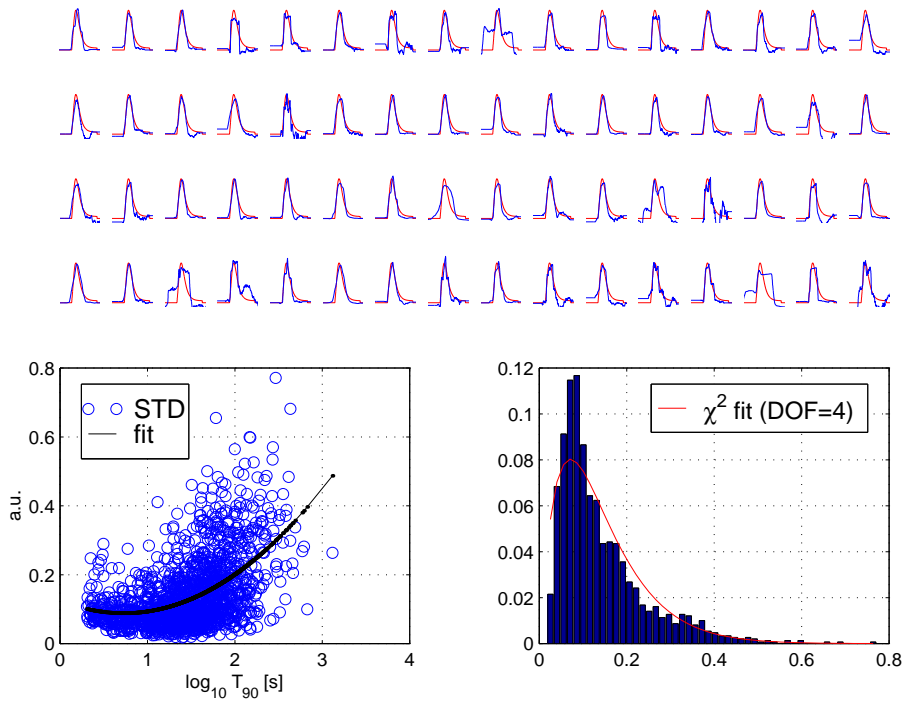}}
\caption{($Top$). Shown is a sample of the nlc$_i$ ($blue$) of the first 32 light curves from Fig. \ref{FIG_match01} by matching to a template $(red)$ using the linear transformations of shift and scaling of the time axis and photon count rates. ($Bottom.$) Shown is the standard deviation (STD) of the deviation between the nlc$_i$ ($i=1,2,\cdots, 1491$) and Template A and its probability density function. The STD increases with duration, resulting from an increasing variability with duration $T_{90}$ with a shallow minimum around $T_{90}=10$ s.}
\label{FIG_match02}    
\end{figure}

We next apply matched filtering to each nlc$_i$ $(i=1,2,\cdots N$) of the BATSE light curves in Fig. \ref{FIG_match01}. It consists of a fit seeking maximal correlation between template and the nlc$_i$ by optimizing Pearson coefficients, in which mean values of photon count rates are ignored. It comprises a three-parameter fit to a template light curve tLC following translation $\Delta t_i$ and scaling $\alpha_i$ of time, as well as a scaling in photon count rate, $\beta_i$, for an optimal fit in the interval $[-1,3]$. This procedure may be seen to be equivalent to an optimal fit in the sense of least squares. Subsequently, the inverse of the transformation of the template thus obtained is applied to the GRB light curve, i.e., a shift $-\Delta t_i$ to the origin and normalization $\alpha_i^{-1}$ in duration and $\beta_i^{-1}$ in count rate. A sample of the resulting nlc$_i$ is shown in Fig. \ref{FIG_match02}. Each match to the template tLC$(t_k)$ may be quantified by the standard deviation $\sigma_i$ of the residuals
\begin{eqnarray}
\rho_i(t_k)=\mbox{nlc}_i(t_k)-\mbox{tLC}(t_k)~~(i=1,2,\cdots, N, ~k=1,2,\cdots, \nu,~t_k\epsilon[-1,3])
\label{EQN_d}
\end{eqnarray}
where $\nu=1600$ in our numerical evaluation. For Template A, the results are shown in Fig. \ref{FIG_match02}.

An a priori estimate of the resolving power of our matched filtering procedure may be derived from the Standard Error in the Mean (SEM). In considering sub-samples of size $N$, such as in moving averages to be considered below, the tandard deviations of the residuals (\ref{EQN_d}) shown in Fig. \ref{FIG_match02}, are
\begin{eqnarray}
SE = \frac{\bar{\sigma}}{\sqrt{N}} \simeq 1.8 \sqrt{\frac{60}{N}} \%.
\label{EQN_SE}
\end{eqnarray}
We next set out to calculate the nLC obtained from various model templates.

\section{Extracting a normalized light curve (nLC) of long GRBs}

Since the nlc$_i$ are normalized, they are bounded on [-1,3]. The nlc$_i$ can now be averaged to obtain a normalized light curve (nLC) representative of {\em all} light curves of long GRBs in the BATSE catalogue:
\begin{eqnarray}
\mbox{nLC}(t_k) = \frac{1}{N}\Sigma_{i=1}^N \mbox{nlc}_i(t_k)
\label{EQN_s0}
\end{eqnarray}
for the ensemble average of the nlc$_i(t_k)$ with standard deviation $\sigma(t_k)$ at each discrete time $t_k=-1+4k/\nu\epsilon[-1,3]$ $(k=1,2,\cdots\nu)$. For a final comparison between the nLC and the template, the nLC is given the same mean value in normalized count rate as the template. 

\begin{center}
\begin{figure}[ht]
\centerline{\includegraphics[angle=00,scale=.45]{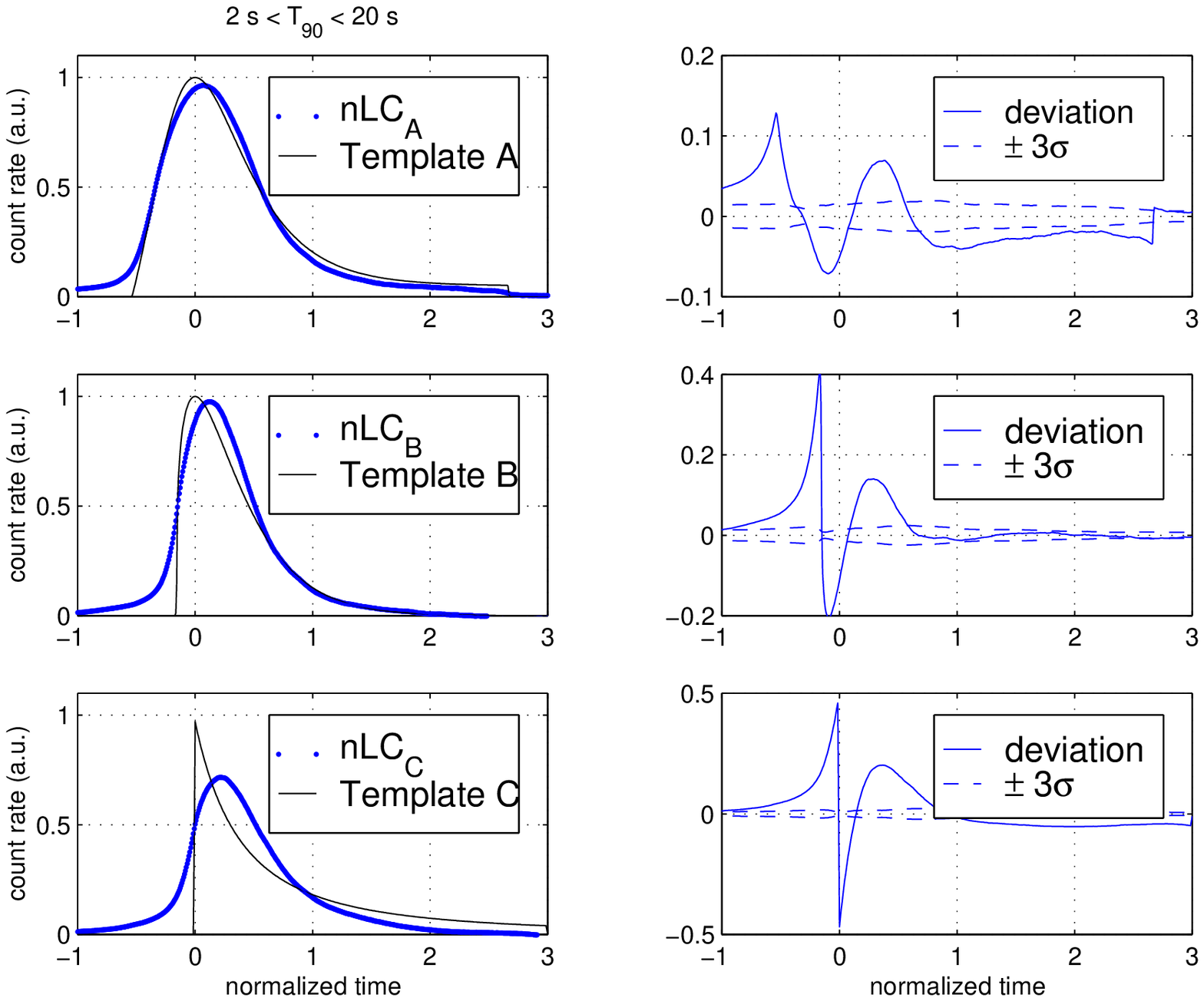}\includegraphics[angle=00,scale=.45]{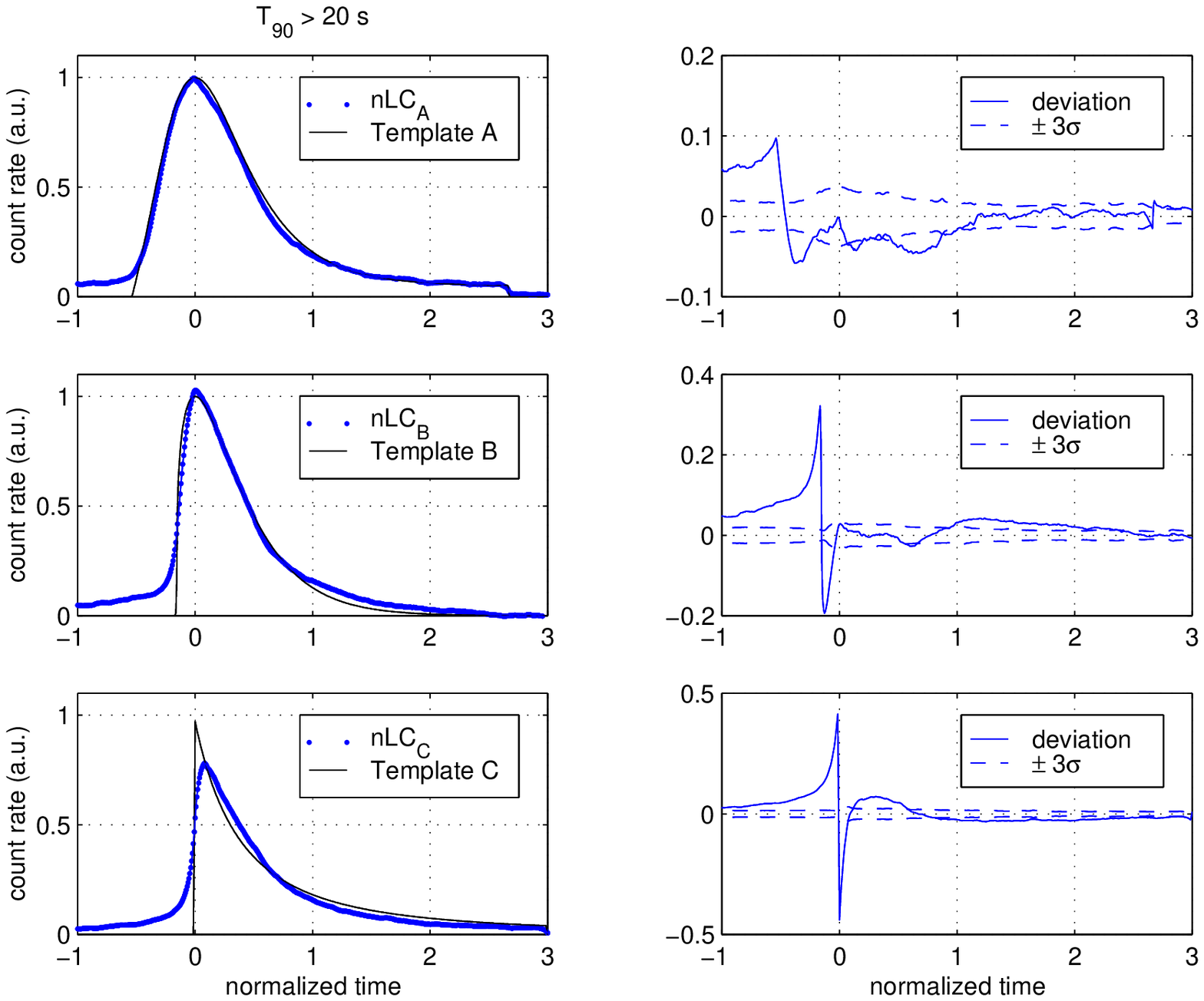}}
\caption{Shown are the nLC $(circles$) generated by model templates A-C ($lines$) for the ensemble of 531 long duration bursts with 2 s $<T_{90}<$ 20 s ($left$) and the ensemble of 960 long bursts with $T_{90}>$20 s $(right)$ and the associated deviations for templates A-C. Here, the standard deviation $\sigma$ is calculated from the square root of the variance of the photon count rates in the ensemble of individually normalized light curves as a function of normalized time.}
\label{FIG_match1}
\end{figure}
\end{center}

\begin{center}
\begin{figure}[ht]
\centerline{\includegraphics[angle=00,scale=.75]{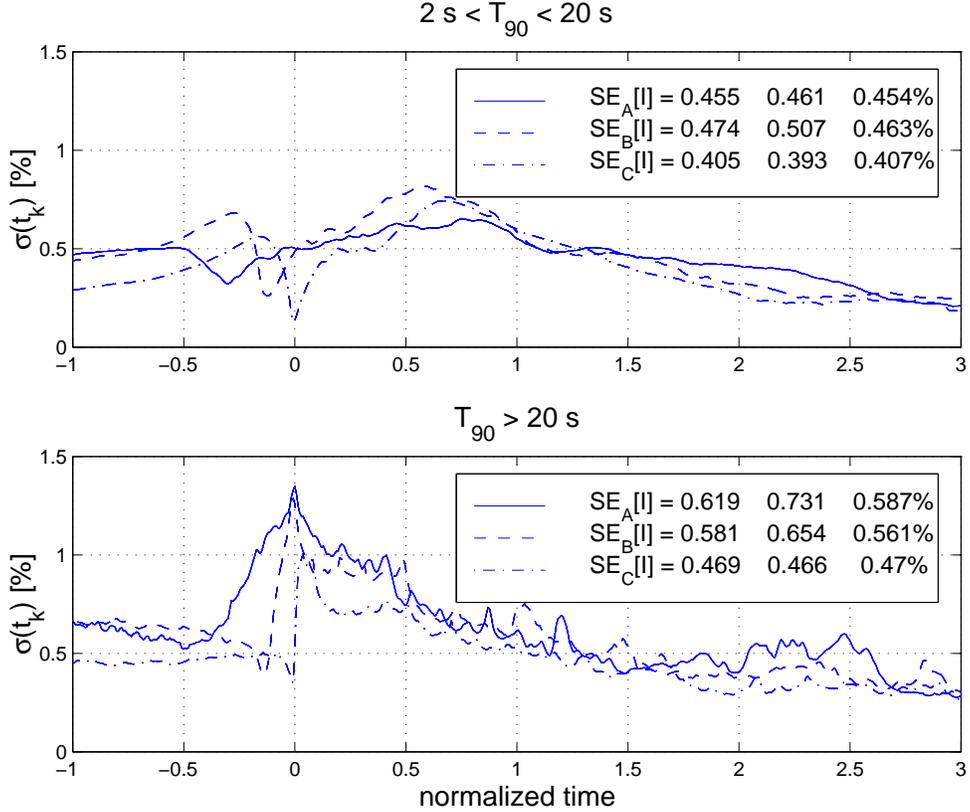}}
\caption{Shown is the degree of convergence in the nLC expressed by the standard error in the mean SE$_i$ in (\ref{EQN_s0}) with $i=$A-C. The three numbers refer to the interval $I=[-1,3]$, the interval $I=[-1,0]$ capturing the rise and the interval $I=[0,3]$ capturing the slow decay post-maximum. Because of the large size of the BATSE catalogue, the results are on average well below 1\%.}
\label{FIG_SE}
\end{figure}
\end{center}

\begin{table}[ht]
{\bf TABLE I.} Summary on the nLC over $t_k\epsilon[-1,3]$ $(k=1,2,\cdots,\nu=1600$) of two subsamples of long GRBs to templates derived from spindown of black holes (\ref{EQN_MA}) and PNS (\ref{EQN_PNS}) shown in Fig. \ref{FIG_match1}. \\ \\
\centerline{
\begin{tabular}{lccccc}
\hline
Quantity & Symbol 	& $2\mbox{~s}<T_{90}<20\mbox{~s}$ & $T_{90}>20\mbox{~s}$ \\
\hline
\hline
mean duration		& $T_{90}$[s]			& 9.7   					& 74.2  			\\
ensemble size		& $N$				& 531					& 960			\\
\\	
(sub-)interval   		& $I$					& [-1,3],~[-1,0],~[0,3]			& [-1,3],~[-1,0],~[0,3] \\
\\
convergence $\sigma$ 	& $SE_A [\%]$     & 0.455,~0.461,~454 		& 0.619,~0.731,~0.587 \\
					& $SE_B [\%]$	 & 0.474,~0.507,~0.463		& 0.0.581,~0.654,~0.561\\
					& $SE_C [\%]$     & 0.405,~0.393,~0.407             & 0.469,~0.466,~0.470\\
\\
deviation $\delta$	&$\delta_A [\%]$            	 & 4.0,~5.6,~2.9   			& 3.2,~5.0,~1.8	\\
				&$\delta_B [\%]$                  & 7.3,~12,~4.2				& 5.1,~8.8,~1.7	\\
				&$\delta_C [\%]$                  & 10.4,~10.0,~9.3			& 6.6,~7.0,~4.7	\\
\\
goodness-of-fit $\chi_{red}^2$	&$\chi_A^2 [\%]$     & 69,~13,~35   			& 27,~65,~8.4	\\
				      		&$\chi_B^2 [\%]$     & 252,~860,~48			& 93,~295,~10.7\\
						&$\chi_C^2 [\%]$     & 1278,~1222,~1050		& 183,~278,~54\\
\hline
\end{tabular}
\label{TABLE_1}}
\mbox{}\hskip0.01in
\end{table}

In this process, baseline levels in background gamma-ray count rates are suppressed by matching templates to data up to an arbitrary offset in count rate. By design, our procedure filters out short and intermediate timescale fluctuations or modulations, that may result from intermittencies in accretion, the torus, shocks and turbulence in the outflow, leaving only the persistent evolution on a secular time scale of evolution of the black hole in each event. By the central limit theorem, (\ref{EQN_s0}) hereby has a well-defined limit as $N$ approaches infinity with asymptotically Gaussian behavior in fluctuations in the process of convergence. 

The nLC generated by the templates tLC of models A-C are shown in the left columns of Figs. \ref{FIG_match1}. To quantify the uncertainty in the nLC resulting from the large but finite sample average, Table 1 lists the mean standard deviation $\sigma(t_k)$ over $t_k=[-1,3]$ in the subsamples of 531 individually normalized light curves nlc$_i(t_k)$ of the two groups 2 s $< T_{90} < $ 20 s and $T_{90}> $ 20 s. It expresses the degree of convergence in (\ref{EQN_s0}) point wise, for each $t_k$. 
{The results in Fig. \ref{FIG_SE} show strong convergence in the nLC($t_k$) at each $t_k$ in the sense of a small standard error in the mean SE$_E\sim 0.5$\%, which is less than the residual deviation $\delta$ between the nLC and the template at hand as listed in Table I.  The resulting $\chi_{red}$ in (\ref{EQN_chi}) hereby effectively reduces to the ratio $\sim \delta/$SE$_i$, measuring a residual error in the templates relative to the uncertainty in the observation. According to Table I, the residual errors are a few percent for Template A, which is more than adequate to distinguish it from various model alternatives.}

\begin{center}
\begin{figure}[ht]
\centerline{\includegraphics[scale=.45]{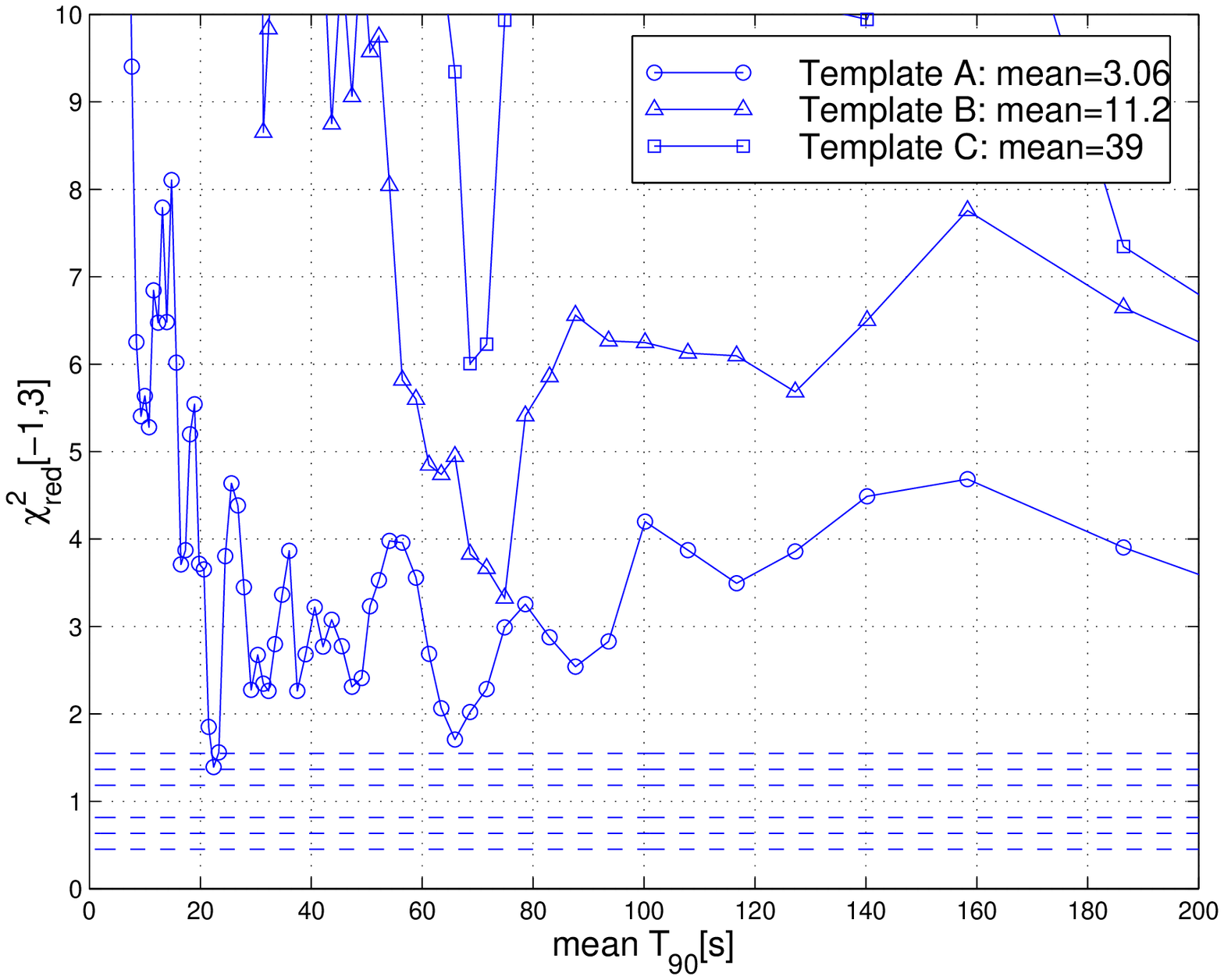}\includegraphics[scale=.45]{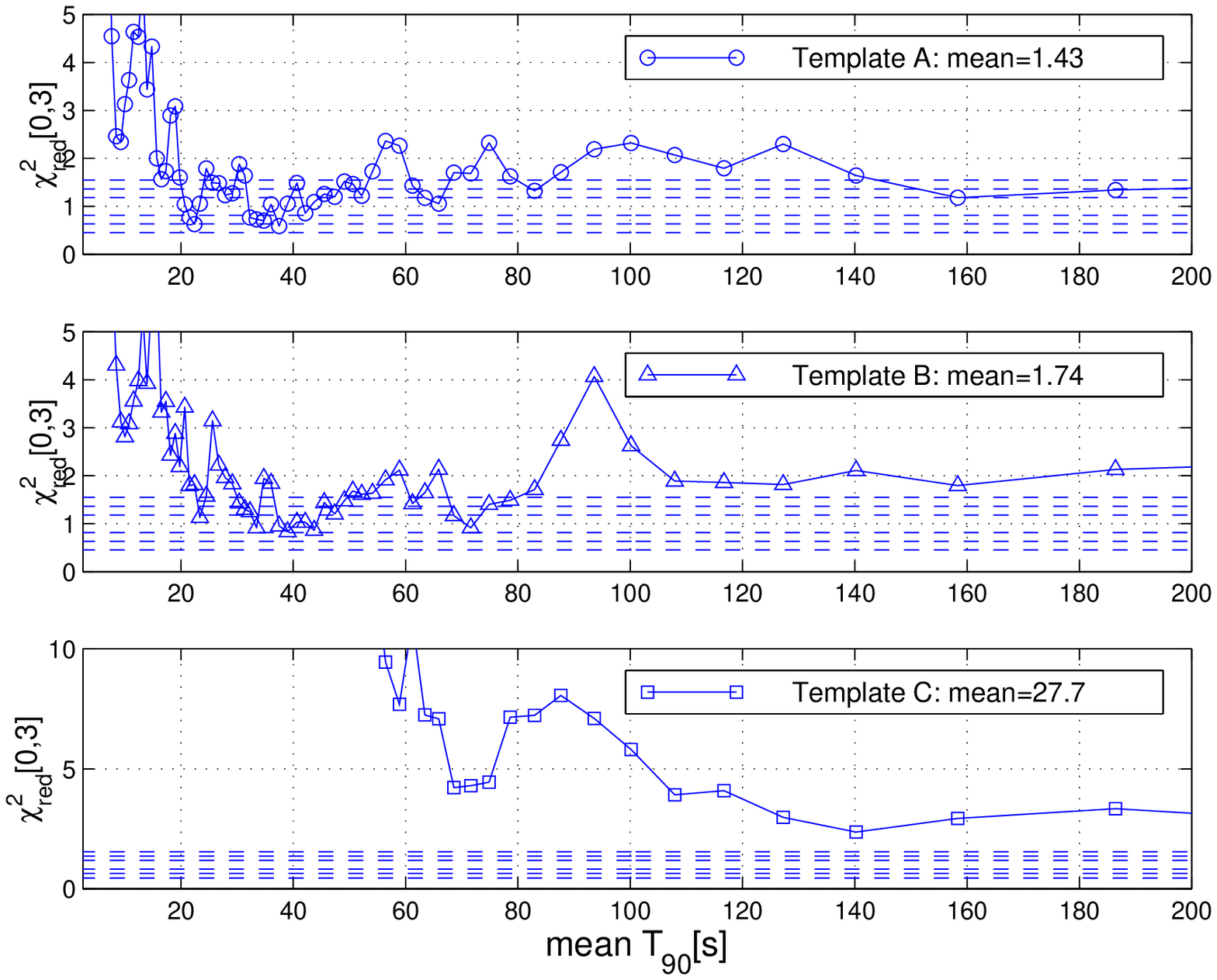}}
\caption{($Left.$) Shown are the $\chi^2_{red}[-1,3]$ for the nLC of Templates A-C as a function of mean durations $T_{90}$ of ensembles of 60 light curves with mean durations in the range $T_{90}>2$ s. There is a noticeable transition in the match for Template A around $T_{90}\simeq 20$ s, which corresponds to a de-redshifted duration of about 10 s. The mean value of the $\chi_{red}^2$ over 20 s $< T_{90} <$ 200 s are given in the legend. Overall, the discrepancy between the nLC and its generating template is smallest for model A. The largest contribution to $\chi_{red}^2[-1,3]$ stems from the onset in the interval $[-1,0]$. (The dashed lines refer to $2\sigma$, 3$\sigma$ and $4\sigma$ intervals of confidence.)
($Right.$) Shown are the $\chi^2[0,3]$ for the prolonged phase of spindown post-maximum in the nLC in the interval $[0,3]$ of normalized time, which is parameter free for models A and C. The mean values 1.43 and 1.74 of the $\chi_{red}^2[0,3]$ for Models A and B, respectively, represent a $2.35\sigma$ and $4.05\sigma$ deviation from 1. Spindown of a PNS (Model C) is essentially ruled out by $\chi_{red}^2\ge 2.32$ for all $T_{90}$ (a $12.7\sigma$ departure from 1). (The dashed lines refer to $2\sigma$, 3$\sigma$ and $4\sigma$ intervals of confidence.)}
\label{FIG_matchC}
\end{figure}
\end{center}

Table 1 lists the deviation between the nLC and its generating template,
\begin{eqnarray}
\delta(t_k)=\mbox{nLC}(t_k)-\mbox{tLC}(t_k).
\label{EQN_d}
\end{eqnarray}
expressed in $\Sigma[I] = \mbox{STD}\left(\left\{\delta(t_k)\right\}_{t_k\epsilon I}\right)$ over the intervals $I=[-1,3]$, $I=[-1, 0]$ during the rise and $I=[0,3]$ during the fall post-maximum in the nLC, and the goodness-of-fit in the match between the nLC to the template tLC by
\begin{eqnarray}
\chi_{red}[I]= \left\{ \frac{N}{\left| I \right|} \Sigma_{t_k\epsilon I}\left[\frac{\mbox{nLC}(t_k)-\mbox{tLC}(t_k)}{\sigma(t_k)}\right]^2\right\}^\frac{1}{2},
\label{EQN_chi}
\end{eqnarray}
where $\sigma(t_k)$ is defined by the SE$_i(t_k)$ shown in Fig. (\ref{FIG_SE}). 

The results unambiguously show {\em three} trends in $\chi_{red}^2$, pointing towards Template A, a prolonged decay in the nLC post-maximum on the interval $I=[0,3]$ and the sub-group of events with durations $T_{90}>$20 s.  Here, Template A is favored with a residual deviation $\delta_A=1.8\%$, whose $\chi_A^2[0,3]=8.4$ is significantly smaller than the $\chi_A^2[0,3]$ of Templates B and C. The residual deviation in Template A of a few percent is well beyond the intended accuracy of our model light curves. (More on this in the subsequent section.) 

The exceptionally large size of the BATSE catalogue allows us to elaborate in some detail on $\chi_{red}^2$ as a function of duration $T_{90}$. Fig. \ref{FIG_matchC} shows $\chi^2$ for a moving average of ensembles of 60 light curves as a function of their mean $T_{90}$. The results for template A show a clear break in convergence and in $\chi^2_{red}$ at $T_{90}\simeq 20$ s, or about 10 s in the comoving frame of the events based on their general correlation to the cosmic star formation rate. Quite generally, a number of processes may be present with time constants up to seconds in all long bursts, that are independent of the durations. Imprints of these processes in the nLC diminishes with increasing mean $T_{90}$ of the ensemble at hand. 
The $n\sigma$ confidence interval for $\chi^2_{red}$ associated with our moving average in time $T_{90}$ over subsamples of nlc$_{i_k}$ $(i_k=1,2,\cdots 60)$ are (e.g. \cite{and10})
\begin{eqnarray}
0.6349\le \chi_{red}^2\le 1.3651~~(n=2),\\
0.4487\le \chi_{red}^2\le 1.5477~~(n=3),\\
0.2693\le \chi_{red}^2\le 1.7303~~(n=4),
\label{EQN_chir}
\end{eqnarray}
represented by the dashed lines in Fig. \ref{FIG_matchC}. Fig. \ref{FIG_matchC} further zooms in on the prolonged spindown phase in the nLC, post-maximum in the normalized time interval $I=[0,3]$ as a function of $T_{90}$. Here, $\chi_{red}^2$ of Model A is within the $3\sigma$ confidence interval, while Model B is outside 4$\sigma$ and Model C is outside the $12\sigma$ confidence level. 

\section{Excess emission in gravitational-waves following a Hopf bifurcation}

{The relatively small deviation between Template A and nLC$_A$ of a few percent points to a sustained presence of an inner disk or torus that extends down to the ISCO for the duration of rapid spin of the black hole.  In the presence of prodigious input from the black hole,  this state requires cooling beyond that facilitated by magnetic winds and MeV neutrinos alone, as the following shows.}

An upper bound $L_w^*$ on the luminosity in disk winds can be derived in the idealized limit in which all angular momentum transport $T_H$ in (\ref{EQN_MT}) is radiated away to infinity by magnetic winds with vanishing emissions in MeV neutrinos and gravitational waves. We can estimate $L_w^*$ by taking a weighted average of the local luminosity $dL_w=\Omega_T^2 dA_\phi^2$ associated with an electromagnetic vector potential $A_a$ supporting a net poloidal flux $\Phi=2\pi dA_\phi$ in a strip $2\pi r dr$ (cf. \cite{van01a}),
\begin{eqnarray}
L_w^*=\kappa \frac{\int_z^{\infty} \Omega_T^2 dE_k }{\int_z^\infty dE_k },
\label{EQN_Lw}
\end{eqnarray}
assuming a constant ratio $dA_\phi^2/dE_k$ of poloidal magnetic field energy-to-kinetic energy in the disk. Here, the disk extends from $z=r_{ISCO}/M$ to infinity. The associated energies follow by integration, $E_H=\int_0^\infty L_H dt$, $E_w^*=\int_0^\infty L_w^*dt$ following a prescription for the distribution $dE_k$. A leading order approximation to $dE_k=2\pi \rho H r dr$ is defined by the Shakura-Sunyaev solution \citep{sha73} for a disk with radial density distribution, $\rho$, scale height, $H$, and perturbed angular velocity, $\Omega_D$, satisfying
\begin{eqnarray}
\rho=\rho_0 \left(\frac{r}{r_0}\right)^{-\frac{15}{8}},~~
H=H_0\left(\frac{r}{r_0}\right)^\frac{9}{8},~~ 
\Omega_D\sim \Omega_T\left(\frac{r_0}{r}\right)^q,
\label{EQN_sha}
\end{eqnarray}
where the $\rho_0$, $H_0$ and $\Omega_T$ denote respective scale factors. Resulting from a positive torque acting on the inner face provided by the central rapidly rotating black hole, we here have $q>\frac{3}{2}$. In this event, we have 
\begin{eqnarray}
E_k = \int_{r_0}^\infty dE_k = \pi \Omega_T^2\rho_0H_0r_0^4 \int_{1}^\infty u^{\frac{9}{4}-2q} du = \frac{\pi \Omega_T^2\rho_0H_0r_0^4}{2q-\frac{13}{4}},
\label{EQN_Ek}
\end{eqnarray}
provided that $q>\frac{13}{8}$ (super-Keplerian motion). In this parameter range, $dE_k$ is a normalizable weight for calculating (\ref{EQN_Lw}) that includes the condition $q>\sqrt{3}$ $(>\frac{13}{8})$ for a sufficiently slender inner disk or torus to be unstable to the non-axisymmetric Papaloizou-Pringle instabilities \citep{pap84,van03}.
\begin{figure}[h]
\centerline{\includegraphics[scale=.6]{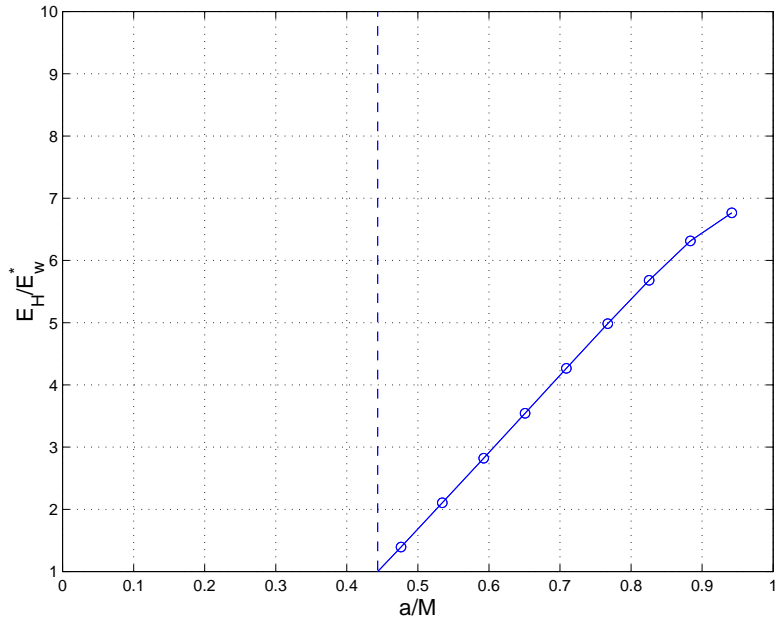}}
\caption{Shown is the ratio of the energy output of the black hole, $E_H$, to the maximum energy in disk winds, $E_w^*$, in the process of spin down as a function of the initial dimensionless spin ratio $a/M$ of the black hole. Beyond the critical value $a/M\simeq 0.4433$, the excess energy $E_H-E_w^*>0$ is generally radiated in a combination of MeV neutrinos and gravitational waves. Observational constraints on the explosion energy in a relativistic core-collapse supernova attributed to disk winds hereby provide an upper bound on the gravitational-wave emissions.}
\label{FIG_Lw}
\end{figure}

In a core-collapse event, the explosion energy ${\cal E}_{SN}$ produced by the momentum of relativistic magnetic winds and any conversion thereof into electromagnetic radiation onto the remnant stellar envelope from within satisfies ${\cal E}_{SN} \simeq \frac{1}{2}\beta E_w$ by momentum conservation, where $\beta$ denotes the velocity of the ejecta relative to the velocity of light. As a result, we have \citep{van11b}
\begin{eqnarray}
E_{gw}+E_{\nu}=E_H-E_w \ge E_H-E_w^* = \beta^{-1} \left(\frac{E_H}{E_w^*}\right){\cal E}_{SN},
\label{EQN_E1}
\end{eqnarray}
where $E_H/E_w^*$ is obtained by numerical integration shown in Fig. \ref{FIG_Lw}, relevant to the case of initial dimensionless spin values $a/M\ge 0.4433$ for which $E_H-E_w^*$ is positive.

The excess energy $E_H-E_w$ in (\ref{EQN_E1}) will be radiated away in an output $E_\nu$ in MeV neutrinos and and output $E_{gw}$ in gravitational waves. While the first is generic, the second may be seen to occur above a critical temperature. 

We recall that for infinitely slender tori, \cite{pap84} derived an index of rotation $q$ for the formation of non-axisymmetric instabilities in torus with radial angular velocity distribution $\Omega(r)$ modeled by a power law, satisfying
\begin{eqnarray}
q>\sqrt{3}, ~~\Omega(r)=\Omega_R\left(\frac{R}{r}\right)^{-q}.
\end{eqnarray}
These instabilities open up angular momentum transport outwards by coupled wave-motion on the inner and outer face of the torus, and simultaneously so for all non-axisymmetric modes $m>1$. However, for astrophysical applications (e.g. \cite{zur89}), we require an extension to wide tori beyond the singular limit of infinitely slenderness, described by finite ratios $0<\delta=b/R<1$ of the minor radius $b$ to the major radius $R$. Instabilities now occur for $q>q_c(\delta,m)\ge\sqrt{3}$ for increasingly many $m$ as the torus becomes more slender, recovering the bifurcation point $q=\sqrt{3}$ in the limit as $\delta$ approaches zero. Fig. 2 shows the neutral stability curves as a function of $\delta$ for various $m$, a quadratic approximation to which for $\sqrt{3}\le q_c\le2$ are \cite{van02}
\begin{eqnarray}
q_c(\delta,m)=\left\{
\begin{array}{llll}
1.73 + 0.27 \left(\frac{\delta}{0.7506}\right)^2 & (m=1),~~
1.73 + 0.27 \left(\frac{\delta}{0.3260}\right)^2 & (m=2)\\
1.73 + 0.27 \left(\frac{\delta}{0.2037}\right)^2 & (m=3),~~
1.73 + 0.27 \left(\frac{\delta}{0.1473}\right)^2 & (m=4)\\
1.73 + 0.27 \left(\frac{\delta}{0.1152}\right)^2 & (m=5),~~
1.73 + 0.27 m^2\left(\frac{\delta}{0.56}\right)^2 & (m\ge 6)
\end{array}
\right.
\label{EQN_Q1}
\end{eqnarray}
showing that instabilities occur for $q>\sqrt{3}$ for increasingly many $m$ with decreasing slenderness $\delta$. 

Heating produces thermal pressure (in addition to magnetic pressures that may be present), which increases the rotation index, $\Omega_T(r)=\Omega_R\left({R}/{r}\right)^q,$ $\Omega_R^2={M}/{R^3},$ about a point mass $M$, to \citep{van03}
\begin{eqnarray}
q=1.5+0.15\left(\frac{R}{4M}\right)\left(\frac{\delta}{0.2}\right)^{-2} T_{10},~~T_{10}\simeq 2L_{\nu,52}^{1/6}\left(\frac{M_T}{0.1M_\odot}\right)^{-1/6},
\label{EQN_q}
\end{eqnarray}
associated with a ratio $\delta = \frac{b}{R}$ of the minor-to-major radius of the torus at temperatures $T=T_{10}10^{10}$ K. At 2 MeV we note comparable Alfv\'en and sound wave velocities in the presence of super strong magnetic fields \citep{van09}. Critical values for the Papaloizou-Pringle instabilities of the first two azimuthal modes $m=1,2$ are \citep{van02,van03}
\begin{eqnarray}
q_{c1}(\delta)=1.73+0.27\left(\frac{\delta}{0.7506}\right)^2,~~q_{c2}(\delta)=1.73+0.27\left(\frac{\delta}{0.3260}\right)^2,
\label{EQN_q12}
\end{eqnarray}
where instability sets in for $q>q_c(\delta,m)$. The value of $\delta$ in (\ref{EQN_q12}) is assumed, and generally may depend on a variety of factors including the history of formation of the torus.  As both thermal and magnetic pressures contribute to (\ref{EQN_q}), it follows that a torus in suspended accretion is susceptible to {\em pressure induced} instabilities. Those that are non-axisymmetric strengthen by gravitational-radiation  back reaction \citep{van02}, whose luminosity is determined by the nonlinear saturation amplitude for unstable non-axisymmetric modes in the torus. An analytical estimate for the low-$m$ multipole mass moments can be given for a flat infrared spectrum in MHD turbulence at the threshold of magnetic stability. The low-$m$ mass inhomogeneities ${\delta m}/{M_T}$ will reach approximately the stability limit ${{\cal E}_B}/{{\cal E}_k}\simeq {1}/{15}$ \citep{van03}, and produce broad line and quasi-periodic emissions in gravitational radiation.

The onset of non-axisymmetric instabilities in the torus for a critical energy input from the black hole (Fig. \ref{FIG_hopf}) is reminiscent of a Hopf bifurcation (e.g. \cite{kel86}), here leading to a sustained emission in gravitational waves for the lifetime of black hole spin. We here identify it with the prolonged spindown identified in the nLC of the GRB light curves of \S6.
\begin{center}
\begin{figure}[ht]
\centerline{\includegraphics[scale=.24]{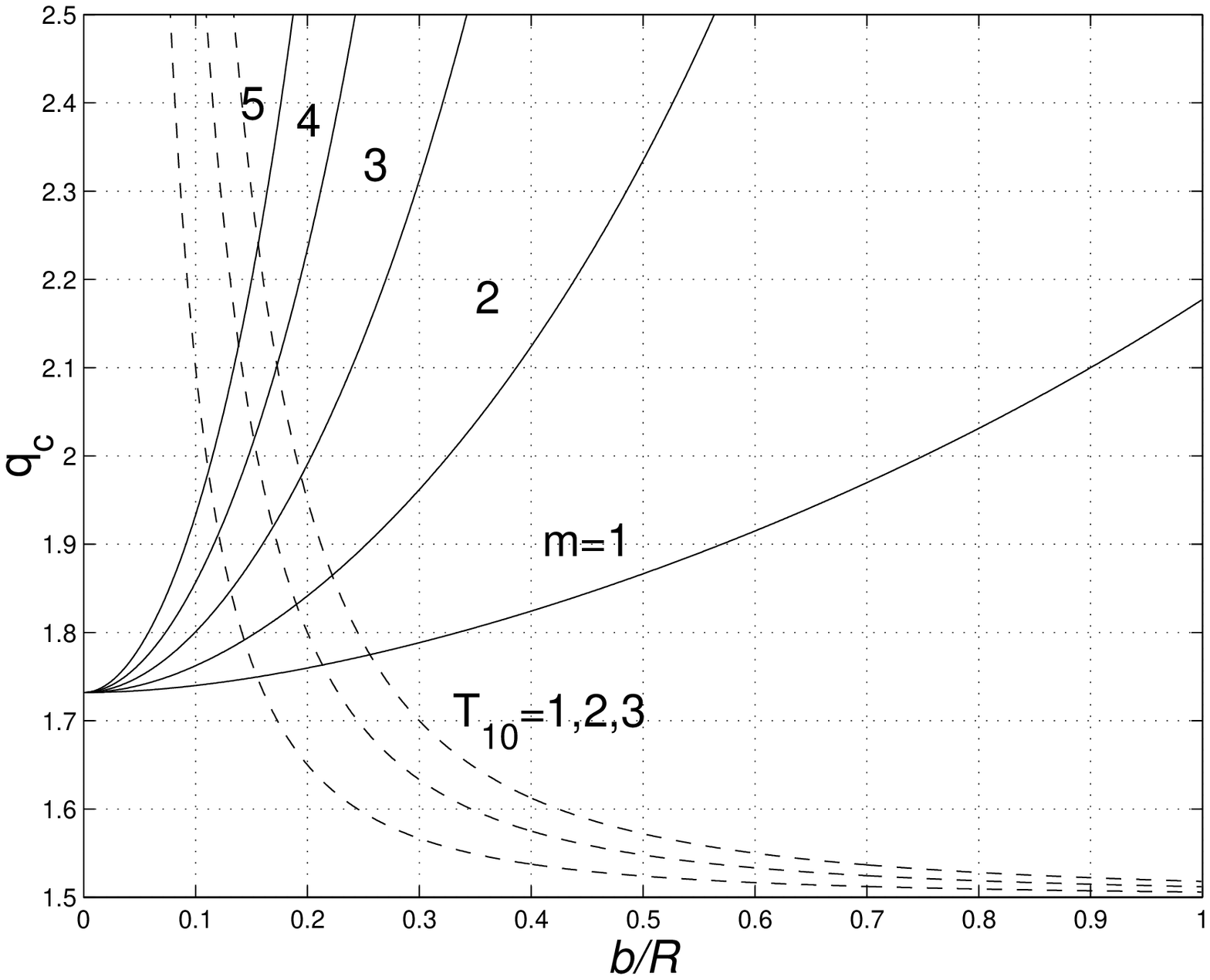}\includegraphics[angle=00,scale=.24]{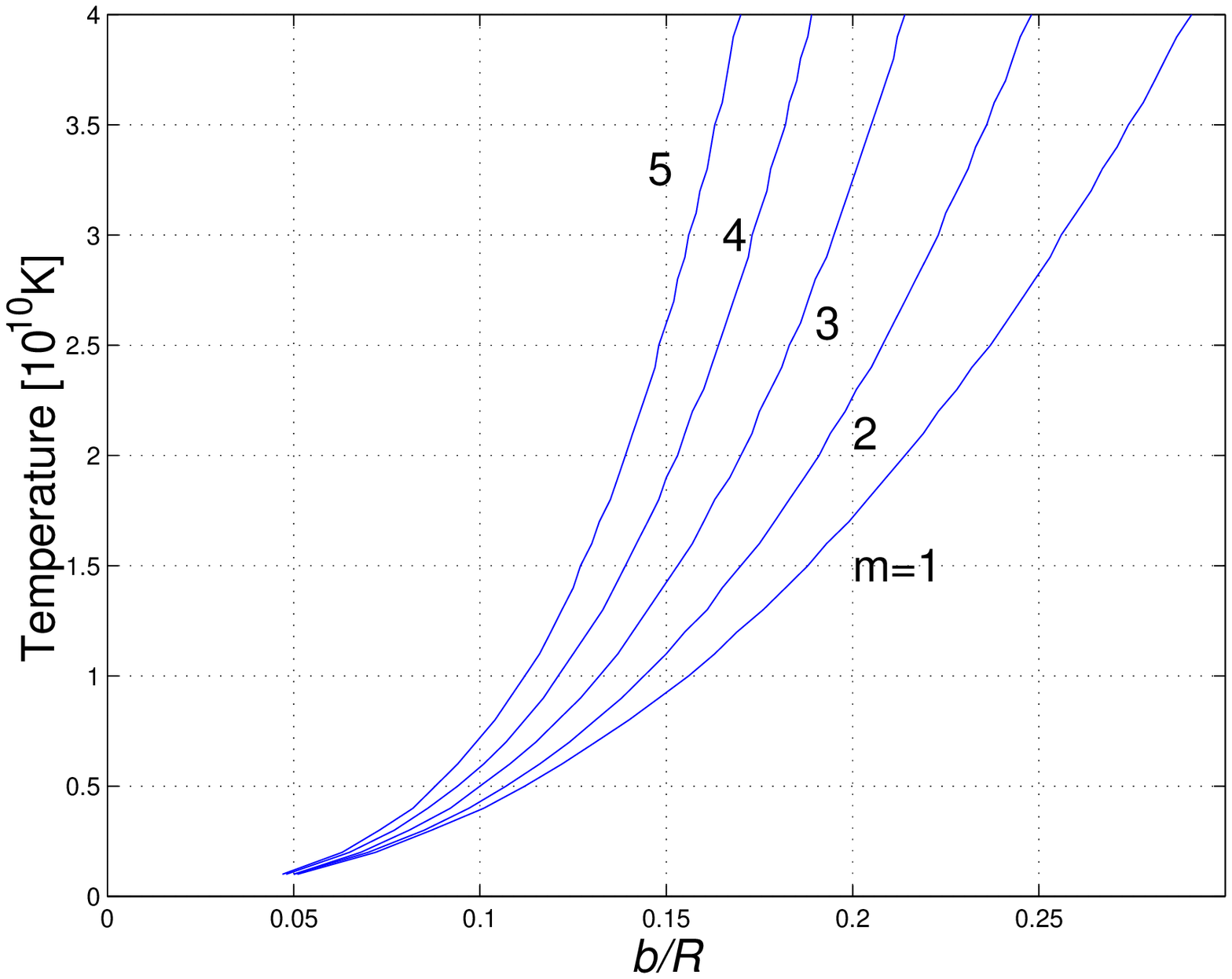}
\includegraphics[scale=.24]{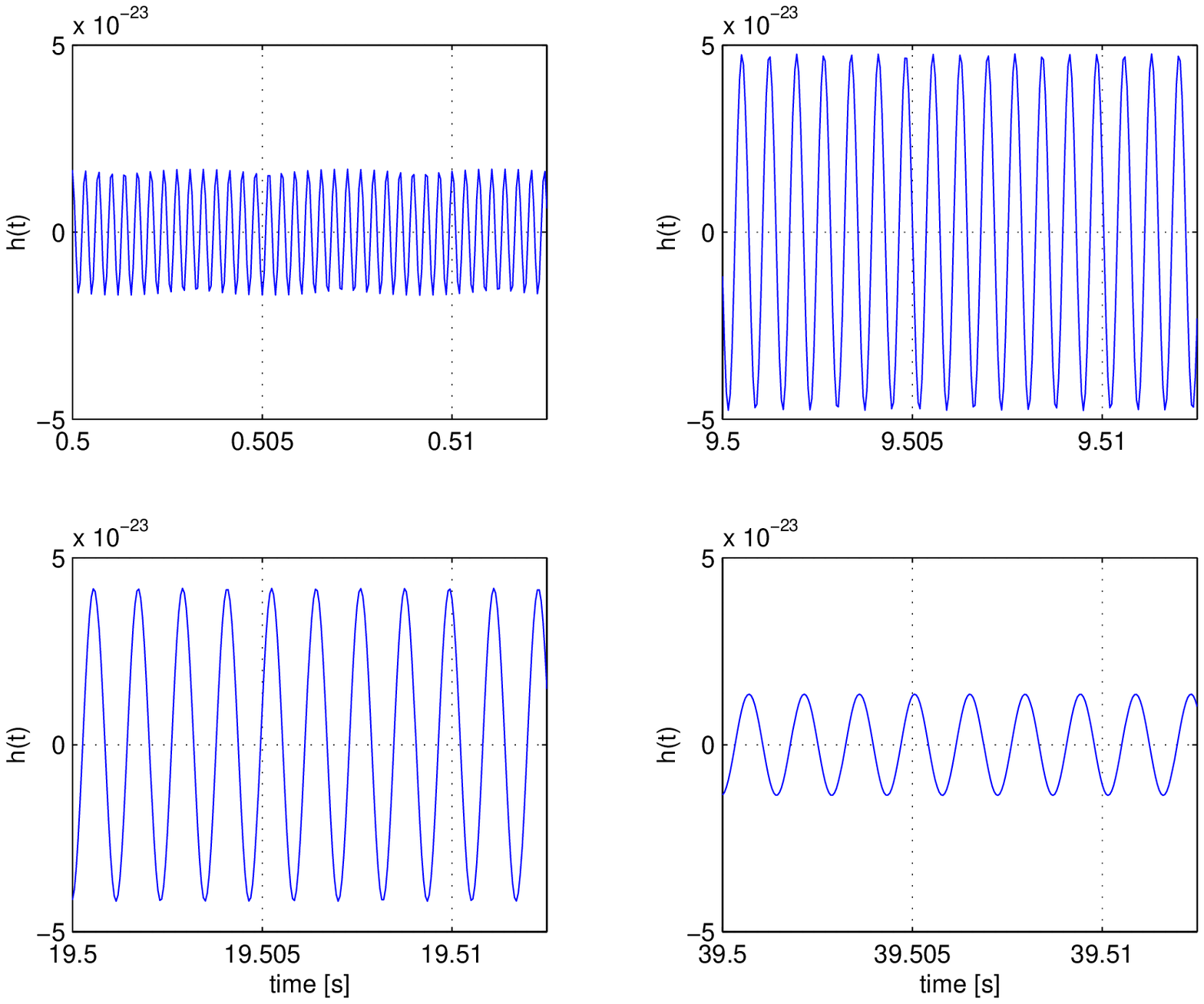}\includegraphics[angle=00,scale=.24]{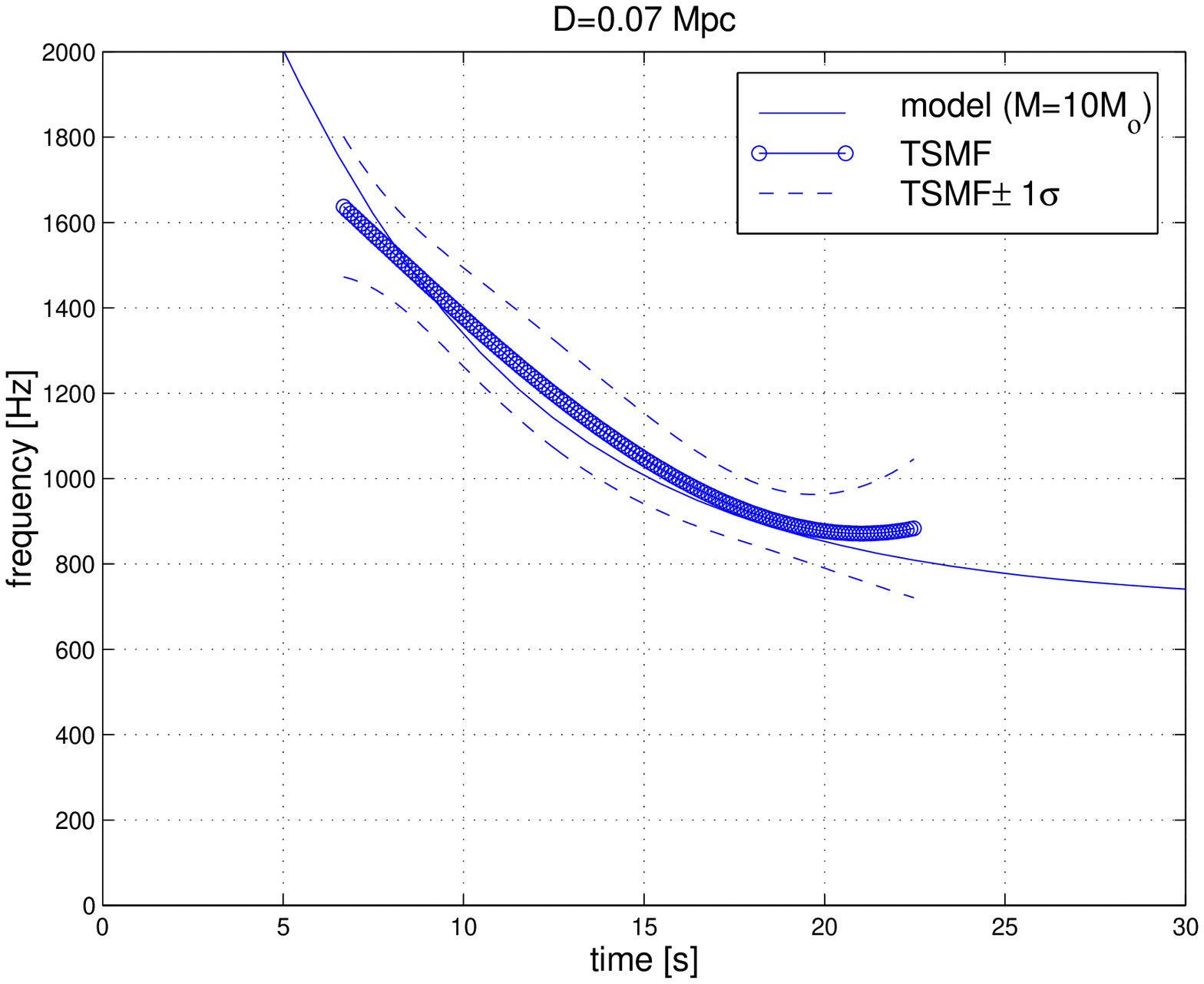}}
\caption{
($Left.$) Diagram showing the neutral stability curves ($solid$ $lines$) for the critical rotation index $q_c$ of buckling modes in a torus of incompressible fluid, which is an extension of the Papaloizou-Pringle instability to arbitrary ratios of minor-to-major radius $b/R$. Curves of $q_c$ are labeled with azimuthal quantum numbers $m=1, 2,\cdots$, where instability sets in above and stability sets in below. In $q \le 2$, the $m=0$ mode is Rayleigh-stable. For $q\equiv 2$, the torus is unstable for $b/R= 0.7385 (m = 1), 0.3225 (m = 2)$ and, asymptotically, for $b/R=0.56/m$ ($m \ge 3$). (Reprinted from \cite{van02}.) Here, the rotation index $q$ as a function of slenderness $b/R$ ($dashed~lines$) is set by the temperature, here shown for $T=T_{10}10^{10}$ K, $T_{10}=1,2,3$ following (\ref{EQN_q}). Heating is provided by the spin energy of the central black hole. The second window shows the critical MeV temperatures for which the index of rotation reaches the critical values $q_c(\delta,m)$ for the onset of non-axisymmetric instability. Generally, higher temperatures are needed for relatively wider tori. This criterion may be relaxed by additional instabilities from magnetic pressures \citep{bro08a}. The third window shows snapshots of the quasi-periodic gravitational wave signal during black hole spindown, scaled to a fiducial duration of about one minute and plotted in frequency with a reduction by a factor of 1000 for illustrative purposes. (Reprinted from \cite{van08a}.) ($Right.$) The long duration negative chirp can be searched for using a dedicated time-sliced matched filtering procedure, here illustrated following signal injection in TAMA 300 noise data. It points to a sensitivity distance of about 35 Mpc for the upcoming advanced detectors (Reprinted from \cite{van11}.)}
\end{figure}\label{FIG_hopf}
\end{center}

In general, the observational consequences of Models A and B follow from the equations of suspended accretion \citep{van01a,van01b}, describing balance of energy and angular momentum flux from the black hole to the inner face of the torus (+) and that emitted to infinity (-) according to
\begin{eqnarray}
\tau_+ = \tau_- + \tau_{GW}+\tau_\nu,~~\Omega_+\tau_+= \Omega_-\tau_-+\Omega_T\tau_{GW}+P_\nu,
\label{EQN_SA1}
\end{eqnarray}
where $\tau_+\propto(\Omega_H-\Omega_+)$, $\tau_{GW}$ and $\tau_-$ denote the angular momentum fluxes in gravitational radiation and in magnetic winds, the latter with $L_{GW}=\Omega_T\tau_{GW}$ and $L_w=\Omega_-\tau_-$. Here, $\Omega_\pm$ denote the angular velocities of the inner and outer faces of the torus. In (\ref{EQN_SA1}), we have neglected energy and angular momentum loss matter outflow, that may be driven by the accompanying MeV neutrino flux. 

For model A, solutions to quasi-periodic emissions in gravitational-wave exist to (\ref{EQN_SA1}) when MHD turbulence has a flat infrared spectrum up to the first geometrical break $m^*\simeq R/b$ (\cite{van01b}, neglecting $\tau_\nu$), wherein luminosities scale with the energy in the magnetic field, and hence the kinetic energy in the torus \citep{van03}. The gravitational wave luminosity is hereby determined self-consistently. At late times, as black hole spin down approaches the fixed point at which its angular velocity approaches that of the ISCO in Model A, we have the gravitational wave frequency \citep{van11}
\begin{eqnarray}
f_{GW}  =  595-704 \mbox{~Hz} \left(\frac{M}{10M_\odot}\right),
\end{eqnarray}
where the frequency window of about 15\% depends on the initial spin of the black hole. It predicts a frequency 1.5-2 k Hz at the end of binary coalescence of two neutron stars and considerably lower frequencies in mergers of neutron stars with a black hole companion or core-collapse in massive stars \citep{van09b}. For model B, we consider (\ref{EQN_SA1}) in the alternative limit with vanishing gravitational radiation. For slender tori described by small $\delta$, $[\Omega]=\Omega_+-\Omega_-\simeq qb/R=2q\delta$, the algebraic solution to (\ref{EQN_SA1}) with $\tau_{GW}=0$ is
\begin{eqnarray}
\frac{\Omega_T}{\Omega_H} = \frac{1}{2}\frac{1}{1+\delta q(\Gamma^2-1)}+O(\delta^2),~~
\frac{P_\nu}{L_w}=2\delta \Gamma^2+O(\delta^2),
\label{EQN_SA2}
\end{eqnarray}where $\Gamma=1/\sqrt{1-v_T^2/c^2}$, assuming a symmetric partition of magnetic flux, in the inner and outer torus magnetosphere extending to the black hole and, respectively, to infinity and. We hereby arrive at Model B, representing a dominant emission in MeV neutrino emissions.

{As the preferred outcome of our matched filtering analysis, Template A points to a {\em self-regulatory} process of formation of multipole mass-moments: energy input from the black hole increases thermal and magnetic pressures which destabilize the inner disk or torus, while gravitational radiation provides cooling which stabilizes by attenuation of the same. This mechanism is characteristic of a Hopf bifurcation \citep{kel86}, describing the onset of stable oscillations at finite amplitude when a control parameter, here temperature, total pressure or rotation index, exceeds a critical value. The result is a stable state producing prolonged emission in gravitational radiation for the lifetime of rapid spin of the black hole.} {The luminosity in gravitational radiation scales with the energy in the poloidal magnetic field (the variance of the magnetic field strength), i.e., the $m=0$ component in the low-$m$ spectrum of moments magnetic fields and mass of the inner disk or torus, that determines the spectrum in gravitational radiation. In the state of forced MHD turbulence in the inner disk or torus as provided by the input from the black hole, they are linked to dissipation at small scales that defines heating and radiation in MeV-neutrinos. In the state of forced MHD turbulence of the inner disk or torus, gravitational radiation and the MeV neutrino emission are inextricably linked representing, respectively, the infrared and UV-spectrum of its mass-motions.}

Very similar considerations may apply to spin down from rapidly rotating neutron stars {\em except} for the fact that its energy reservoir in angular momentum is about two orders of magnitude smaller than that of rotating stellar mass black holes expected to form in some of the most relativistic core-collapse supernovae. For this reason, these emissions are not considered here.

\section{Conclusions}

Core-collapse of massive stars produces black holes and PNS, where the apsherical relativistic events point to a rotationally powered inner engine. Some of these events are hyper-energetic, that can not all be attributed to spindown of a PNS. The association of {\em some but not all} long GRBs with core-collapse of massive stars points to inner engines with a hidden time scale of tens of seconds different from accretion. We therefore pursue a search for spindown of a black holes or PNS on the time scale of duration of long GRBs using three different model templates for the light curve of GRBs, defined by the spin of an initially rapidly spinning black hole by surrounding matter at the ISCO (A), further out (B) and spindown of a PNS (C). For each, we extracted a nLC from the BATSE catalogue of long GRBs by application of matched filtering.

{By its large size, the BATSE catalogue allows the extraction of nLC's with an SE of less than one percent. This resolving power is adequate to distinguish various model alternatives considered in the present analysis.} {The match between nLC$_A$ extracted using Template A is within a few percent, whose prolonged decay post-maximum identifies the parameter free process of spindown in the Kerr metric.

We find that nLC$_A$shows a gradual rise in their light curve over about 16\% of their duration according to (\ref{EQN_tau}), that represents the initial phase of spindown of an extremal black hole $(a=M)$ to $a/M=0.8388$ (Fig. \ref{FIG_T}). A $\chi^2_{red}$ analysis shows a best-fit for Template A relative the the alternatives B and C. As a function of duration, the $\chi_{red}^2$ generally improves with $T_{90}$ featuring a pronounced break at $T_{90}\simeq 20$ s in case of nLC$_A$. Zooming in on the prolonged phase of spindown post-maximum in the nLC, the fit falls within the 2.35$\sigma$ confidence interval, that should be contrasted with the $\chi_{red}^2$ of Templates B and C, which fall outside the 4$\sigma$ and, respectively, 12$\sigma$ confidence intervals. Consequently, Template C, representing spindown of a PNS, is essentially ruled out. 

The break in the goodness-of-fit at durations $T_{90}\simeq 20$ s corresponding to a de-redshifted duration of about 10 s in view of the general correlation of long GRBs with the cosmic star formation rate. Beyond, it approximately levels off, providing evidence that their inner engines are {\em normalizable}. This is different from a standard energy reservoir on the basis of a mean true energy output $E_\gamma\sim 10^{51}$ in gamma-rays as originally envisioned by \cite{fra01}, in allowing for a diversity in the true energy in GRB-afterglow emissions. For instance, the brightest {\em Swift} events include GRB 050820A with a total energy output of $E_{tot}\sim 4.2\times 10^{52}$ erg, i.e., $E_\gamma$ and kinetic energies inferred from its afterglow  \citep{cen10}. 

The nLC refers to the shape in the normalized light curve of the inner engine in the limit of averaging a large number of individually normalized light curves. Inspection of individual BATSE light curves reveals bursts with a smooth onset with remarkably good fits to template A \citep{van09}, as well as individual bursts with a prompt onset with remarkably good fits to either template B or C. Our nLC is a statistical result on the ensemble of BATSE bursts that is not intended to address features of specific, individual events or sub-classes of progenitors. Indeed, long GRBs may well derive from both core-collapse events and mergers and, if so, the onset of their respective light curves might be different, where the former is subject to the process of a break-out of jets through the remnant stellar envelope (associated with a time scale of about 10 s, e.g., \cite{bro11,nak11}), whereas the latter is not.

A normalizable inner engine for long GRBs is consistent with having negligible memory on their formation history, especially when formed as near-extremal objects in core-collapse events. The distribution of their durations then depends largely on the poloidal magnetic field strength, that, for the first, further depends on the mass of the black hole and the surrounding disk. Thus, normalization of durations effectively collapses the intrinsic light curves to a unique light curve. Normalization will be effective when the durations are long relative to any of the small to intermediate time scales that are uncorrelated to the duration of the bursts, such as time scales of accompanying (magneto-)hydrodynamic processes in the disk and the inner torus magnetosphere surrounding the black hole, in the evolution of the dynamo in a PNS, or time scales associated with the break-out of baryon-poor jets through a remnant stellar envelope. The former would include, for instance, dynamical time scales and time scales of instabilities, neither of which is expected longer than a few seconds, while the latter is generally expected to be a few times the light crossing time of the progenitor star. 

The contemporaneous formation of ultra-relativistic capillary jets by frame dragging along the spin axis and spindown against surrounding matter are mediated by frame dragging. These processes combined are only beginning to be approached by numerical simulations. Earlier results show that ultra-relativistic jets, sufficient to produce the GRBs that we see, can not be produced without general relativistic effects \citep{nag07}. More recently, a positive correlation of jet formation with black hole spin is found \citep{nag11}, but a successful formation of ultra-relativistic jets awaits further developments. Our model A suggests that this may require pair-creation (which falls outside the realm of ideal MHD) along open magnetic flux tubes that extend from the black hole event horizon to infinity. A numerical simulation of a complete GRB light curve would further require taking into account black hole spindown against the surrounding matter, where the latter develops a state of forced turbulence, instabilities and multimessenger radiation processes. 

GRBs from rotating black holes give an attractive outlook for unification, as they appear naturally in various progenitor scenarios associated with core-collapse supernovae and mergers of neutron stars with another neutron star or companion black hole. Those involving rapidly rotating black holes can account for long GRBs with supernovae in CC events, and without supernovae such as GRB 060614 in mergers events.  Long GRBs are herein spin powered, not accretion powered \citep{woo93,pop99,che07}. The latter would force the black hole to continuously spin up \citep{kum08}, here in contradiction with the observed long duration decay in the nLC over about 84\% of $T_{90}$ post maximum. Mergers producing long GRBs with no supernovae would include binary coalescence of two neutron stars, as all of these produce rapidly rotating black holes \citep{bai08}. Any merger scenario for long GRBs with no supernovae a priori rules out inner engines in the form of rapidly rotating neutron stars. 

Our identification of black hole spindown in the GRB light curves shows that most of the rotational energy is transferred to the surrounding matter, proposed earlier in \cite{van99}, which far exceeds that observed in the true energy output in the GRB-afterglow emissions. The result is a major output in MeV neutrinos and gravitational waves by catalytic conversion in the inner disk or torus, since this energy exceeds that which can be radiated off in magnetic winds. While the MeV neutrino bursts are difficult to detect even from sources in the local Universe, collectively they inevitably contribute to the cosmological background in MeV neutrinos that may be observationally relevant (cf. \cite{nag03}). 

In the process of black hole spindown, we identify a Hopf bifurcation giving rise to sustained non-axisymmetric instabilities in the surrounding matter for the lifetime of black hole spin, in addition to any magnetic-pressure instabilities that may give rise to the same \citep{bro08a}. It lends credence to a quasi-periodic emission in gravitational waves for a duration of tens of seconds that may be detectable out to a distance of 35 Mpc by the advanced ground based gravitational-wave detectors presently in development \citep{van11}. The anticipated output in gravitational radiation that may be dominant over MeV neutrino emissions, promising one step beyond SN1987A, whose $>10$ MeV neutrino burst (compiled in \cite{bur87}) already exceeded the output in all electromagnetic radiation. It is therefore fortuitous that the goodness-of-fit of nLC$_A$ exceeds that of nLC$_B$.

{\bf Acknowledgments.} The author gratefully acknowledges constructive comments from the referee and stimulating discussions with Amir Levinson, Massimo Della Valle and the anonymous referees. The matched filtering procedure has been implemented in Lahey Fotran 95.


\end{document}